\pdfoutput=1  
\documentclass[a4paper,fleqn,usenatbib,useAMS]{mnras}

\usepackage{newtxtext,newtxmath}

\usepackage[T1]{fontenc}
\usepackage{ae,aecompl}

\usepackage{amsmath}
\usepackage{graphicx}
\usepackage{natbib}
\usepackage[noabbrev]{cleveref}
\usepackage{booktabs} 
\usepackage[toc]{appendix}
\usepackage{layouts} 
\usepackage{subcaption} 
\newcommand{\numrepeats}{\texttt{num\_repeats}}
\newcommand{\nlive}{\texttt{nlive}}
\newcommand{\efr}{\texttt{efr}}
\newcommand{\e}{\mathrm{e}}
\newcommand{\Z}{\ensuremath{\mathcal{Z}}}
\renewcommand{\d}[1]{\ensuremath{\operatorname{d}\!{#1}}}
\newcommand{\thcomp}[1]{\theta_{\hat{#1}}}
\newcommand{\thmean}[1]{\overline{\thcomp{#1}}}
\newcommand{\thi}{\thcomp{i}}
\newcommand{\nestcheck}{\href{https://github.com/ejhigson/nestcheck}{\textcolor{black}{\texttt{nestcheck}}}}
\newcommand{\nestcheckurl}{\href{https://github.com/ejhigson/nestcheck}{\url{https://github.com/ejhigson/nestcheck}}}
\newcommand{\dynesty}{\href{https://github.com/joshspeagle/dynesty}{\textcolor{black}{\texttt{dynesty}}}}
\newcommand{\dynestyurl}{\href{https://github.com/joshspeagle/dynesty}{\url{https://github.com/joshspeagle/dynesty}}}
\newcommand{\dyPolyChord}{\href{https://github.com/ejhigson/dyPolyChord}{\textcolor{black}{\texttt{dyPolyChord}}}}

\newcommand{\diagnosticcodeurl}{\href{https://github.com/ejhigson/diagnostic}{\url{https://github.com/ejhigson/diagnostic}}}
\newcommand{\PolyChord}{\texttt{PolyChord}}
\newcommand{\MultiNest}{\texttt{MultiNest}}
\newcommand{\perfectns}{\texttt{perfectns}}

\title[Diagnostic tests for nested sampling]{\nestcheck: diagnostic tests for nested sampling calculations}

\author[E. Higson et al.]{%
Edward Higson,$^{1,2}$\thanks{E-mail: e.higson@mrao.cam.ac.uk}
Will Handley,$^{1,2}$
Michael Hobson$^{1}$
and Anthony Lasenby$^{1,2}$
\\
$^{1}$Astrophysics Group, Battcock Centre, Cavendish Laboratory, JJ Thomson Avenue, Cambridge CB3 0HE, UK\\
$^{2}$Kavli Institute for Cosmology, Madingley Road, Cambridge, CB3 0HA, UK
}

\date{Accepted XXX\@. Received YYY\@; in original form ZZZ}

\pubyear{2018}

\volume{{\rm in press}}
\begin{document}\label{firstpage}
\pagerange{\pageref{firstpage}--\pageref{lastpage}}
\maketitle


\begin{abstract}
Nested sampling is an increasingly popular technique for Bayesian computation, in particular for multimodal, degenerate problems of moderate to high dimensionality.
Without appropriate settings, however, nested sampling software may fail to explore such posteriors correctly; for example producing correlated samples or missing important modes.
This paper introduces new diagnostic tests to assess the reliability both of parameter estimation and evidence calculations using nested sampling software, and demonstrates them empirically.
We present two new diagnostic plots for nested sampling, and give practical advice for nested sampling software users in astronomy and beyond.
Our diagnostic tests and diagrams are implemented in \nestcheck{}: a publicly available\textsuperscript{1} Python package for analysing nested sampling calculations, which is compatible with output from \MultiNest{}, \PolyChord{} and \dyPolyChord.
\end{abstract}



\section{Introduction}
\setcounter{footnote}{1}
\footnotetext{Available at \nestcheckurl{}.}

Nested sampling \citep{Skilling2006} is a method for Bayesian analysis which simultaneously provides Bayesian evidences and posterior samples.
The popular \MultiNest{} \citep{Feroz2008,Feroz2009,Feroz2013} and \PolyChord{} \citep{Handley2015a,Handley2015b} implementations are now used extensively in many areas of science, and in particular in astronomy; see for example \citet{Samushia2014,Joudaki2016,PlanckCollaboration2015,Desvignes2016,DESCollaboration2017,Chua2018}.
Though originally designed for evidence calculation, nested sampling is now widely employed for parameter estimation and performs well compared to Markov chain Monte Carlo (MCMC)-based alternatives for multimodal and degenerate posteriors due to having no thermal transition property.
In addition the \PolyChord{} implementation is designed to handle higher dimensional problems.

Methods for numerically estimating the uncertainty in nested sampling results due to the stochasticity of the nested sampling algorithm are now available for both evidence calculations \citep[see][]{Skilling2006,Keeton2011} and parameter estimation \citep[see][]{Higson2017a}.
However, all of these techniques assume that the nested sampling algorithm was executed perfectly --- which requires sampling randomly from the prior within a hard likelihood constraint.
This can only be done exactly in special cases, such as for spherically symmetric calculations using \perfectns{} \citep{Higson2018perfectns}.
Nested sampling software used for practical problems can only perform such sampling approximately and as a result may produce additional errors --- for example due to correlations between samples, or due to sampling from only part of the prior volume contained within a likelihood constraint.
We term these additional errors {\em implementation-specific effects\/} to distinguish them from the intrinsic stochasticity of the nested sampling algorithm.

Diagnosing whether significant implementation-specific effects are present is of great practical importance for researchers as they can cause large uncertainty in results and lead to potentially incorrect conclusions --- such as, for example, if the calculation misses a significant mode\footnote{Here we refer to cases where the software does not detect the mode and, as a result, samples are not drawn from the entire prior volume within specified likelihood constraints. Another less common problem is that, if the number of live points is very low, a given run might not contain a single sample within a particular mode even when the nested sampling algorithm is performed perfectly; this is not an implementation-specific effect according to our definition.} in a multimodal posterior.
Conversely, if implementation-specific effects are shown to be negligible, users can simply increase the number of live points for more accurate results and can confidently use standard techniques to estimate numerical uncertainty from the nested sampling algorithm.

Typically software has settings which the user can adjust to reduce implementation-specific effects at the cost of increased computation, such as \PolyChord's \numrepeats{} and \MultiNest's \efr{} (see \Cref{sec:practical} for more details).
Assessing if the software is able to explore the posterior reliably is therefore particularly useful when taking significantly more samples is computationally costly, as is often the case for high-dimensional problems.
In the authors' experience, software users typically try to check their results by running a calculation several times and qualitatively assessing if the posterior distributions look similar in each case.
However this is not very reliable and does not differentiate between implementation-specific effects and the expected variation from the inherent stochasticity of the nested sampling algorithm.

We are not aware of any diagnostic tests in the literature for checking calculation results for practical problems for implementation-specific effects, although \citet{Buchner2016a} proposes a diagnostic for evidence calculations which uses analytically solvable test problems.
In contrast Markov chain Monte Carlo (MCMC)-based methods, which do not require sampling within a hard likelihood constraint, have an extensive literature on diagnostics for practical problems \citep[see for example][]{Cowles1996,Hogg2017}.

This paper introduces new heuristic tests and diagrams to check the reliability of nested sampling results for practical problems, and to determine if the software settings should be changed.
It is also intended to serve as a practical guide for nested sampling practitioners based on the authors' experience using nested sampling software.
We begin with a brief overview of the nested sampling algorithm and its associated errors in \Cref{sec:intro}, and discuss the challenges of detecting implementation-specific effects in \Cref{sec:main_idea}.
We then introduce our new diagnostic tests:
\begin{itemize}
    \item \Cref{sec:plots} discusses diagnostic plots and presents two new diagrams for nested sampling (illustrated in \Cref{fig:bs_param_dist,sub:param_logx_gaussian,sub:param_logx_loggamma_mix});
    \item \Cref{sec:implementation_errors} describes how the implementation-specific effects can be measured from a number of nested sampling runs;
    \item \Cref{sec:diagnostics} introduces diagnostic tests which can be applied to pairs of nested sampling runs and are useful when few runs are available.
\end{itemize}
We empirically test the effects of changing nested sampling software settings and the dimension of the problem on both implementation-specific effects and total calculation errors in \Cref{sec:practical}; the tests use \PolyChord{}, although the discussion and conclusions are relevant for other software.
Our practical advice for software users is summarised in~\Cref{sec:advice}.
Finally in \Cref{sec:planck} we apply our methods to astronomical data from the \textit{Planck} survey.
Our diagnostic tests and diagrams are implemented in \nestcheck{} \citep{Higson2018nestcheck}; an open source Python package for analysing nested sampling calculations.
\nestcheck{} is compatible with output from a variety of nested sampling software packages, including \MultiNest{}, \PolyChord{} and \dyPolyChord{} \citep{Higson2018dypolychord}.

\section{Background: nested sampling and sampling errors}\label{sec:intro}

This section provides a brief overview of the nested sampling algorithm and the sampling errors involved in the process --- for more details see \citet{Higson2017a}.
A comparison of nested sampling with other sampling methods is beyond of the scope of this paper; for this we refer the reader to \citet{Allison2014} and \citet{Murray2007}.

\begin{figure}
    \includegraphics[width=\linewidth]{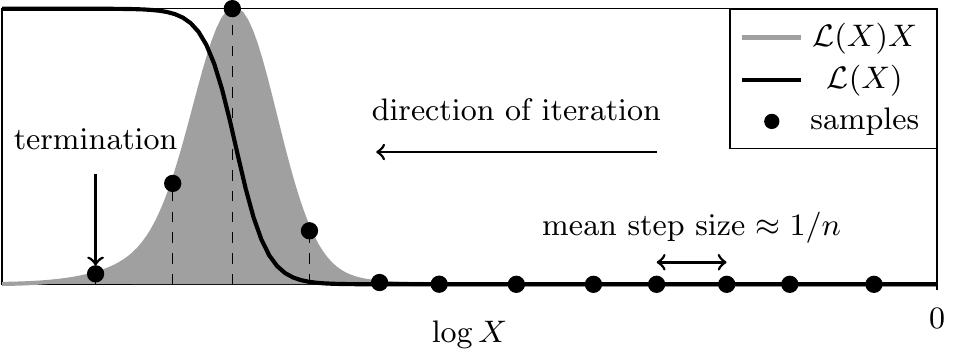}
    \caption{Illustration of nested sampling with a constant number of live points $n$ \citep[reproduced from][]{Higson2017a}.
    The algorithm samples an exponentially shrinking fraction of the prior $X$ as it moves towards increasing likelihoods.
    The relative posterior mass contained at each $\log X$ value is proportional to $\mathcal{L}(X)X$, where $\mathcal{L}(X) \equiv X^{-1}(\mathcal{L})$.}%
    \label{fig:ns_evidence}
\end{figure}

Nested sampling \citep{Skilling2006} performs Bayesian computations by maintaining a set of samples from the prior $\pi(\btheta)$, called {\em live points}, and repeatedly replacing the point with the lowest likelihood $\mathcal{L}(\btheta)$ with another sample from the region of the prior with a higher likelihood.
The samples which have been removed, termed {\em dead points}, are then used for evidence calculations and posterior inferences (the live points remaining when the algorithm terminates can also be included).
The fraction of the prior volume remaining after each point $i$ with likelihood $\mathcal{L}_i$, which is defined as
\begin{equation}
    X(\mathcal{L}_i) \equiv \int_{\mathcal{L}(\btheta ) > \mathcal{L}_i} \pi (\btheta ) \d\btheta,
    \label{equ:X_definition}
\end{equation}
shrinks exponentially; this process is illustrated schematically in~\Cref{fig:ns_evidence}.
The shrinkage at each step is unknown but is estimated statistically and used to weight the samples produced.

The sampling errors from this process can be estimated by dividing a completed nested sampling run with some number of live points into many valid nested sampling runs with only one live point.
These single live point runs, termed {\em threads}, can then be resampled using standard techniques such as the bootstrap as described in Section 4 of \citet{Higson2017a}.
The resampling is valid as the $\log X$ values of the dead points of a nested sampling run with $n$ live points are a Poisson process with rate $n$, so hence the $\log X$ values for the dead points in each of its constituent threads form a Poisson process of rate 1.
Here and in the remainder of this paper $\log$ denotes the natural logarithm.

\section{Measuring implementation-specific effects}\label{sec:main_idea}

This paper is concerned with developing practical diagnostics for assessing whether nested sampling calculation results contain implementation-specific effects due to imperfect execution of the nested sampling algorithm.
It is important to emphasis that diagnosing such effects without additional information about the likelihood and prior is very challenging problem, and it is impossible to conclude {\em a priori\/} with certainty that they are not present.
For example, one cannot eliminate the possibility of missing an extremely narrow mode for a general posterior without an exhaustive search of the parameter space \citep{Wolpert1997}.
\citet[][Section 5]{Hogg2017} provide an interesting and analogous discussion of the similarly heuristic nature of MCMC convergence tests.
In addition, nested sampling's iteration towards successively higher likelihoods means it never reaches a steady state. As a result heuristics based on autocorrelation of samples like those used in testing for MCMC convergence cannot be applied.

The main idea behind the diagnostic tests we present is to assess if the variation of the results of different nested sampling runs is consistent with the statistical properties expected of nested sampling without implementation-specific effects.
Consequently, these diagnostics require multiple nested sampling runs.
A limitation of this approach is that a systematic bias in the calculation results will lead to the implementation-specific effects being underestimated, although they are still likely to be detectable.
Such cases have been observed in the literature for evidence calculations with challenging posteriors \citep[see for example][]{Beaujean2013}; we discuss systematic bias in detail in \Cref{sec:bias}.
Furthermore our diagnostics are unable to detect implementation-specific effects which do not change the variation of the runs, although we have not come across such a case in practice.
A theoretical example would be if every run available missed a significant mode while exploring all the rest of the parameter space correctly.

\subsection{Test problems}\label{sec:test_cases}

We now introduce two test problems, which we will use to demonstrate the diagnostic tests presented in the following sections.

As an example of a simple likelihood, we consider a $d$-dimensional Gaussian with $\sigma=1$ centred on the origin
\begin{equation}
    \mathcal{L}(\btheta) = {(2 \pi)}^{-d/2} \e^{-{|\btheta|}^2 / 2}.
    \label{equ:gaussian}
\end{equation}
We also use the challenging LogGamma-Gaussian mixture model likelihood introduced by \citet{Beaujean2013}, which was designed to represent a particle physics problem involving heavy-tailed distributions and several distinct modes.
In this case $\mathcal{L}(\btheta) = \prod_{i=1}^d \mathcal{L}(\thi)$ with
\begin{equation}
\begin{split}
  \mathcal{L}(\thcomp{1}) &= \frac{1}{2} \mathrm{LogGamma}(\thcomp{1} - 10|1, 1) + \frac{1}{2} \mathrm{LogGamma}(\thcomp{1} + 10|1, 1),\\
  \mathcal{L}(\thcomp{2}) &= \frac{1}{2} \mathrm{Normal}(\thcomp{2} - 10|0, 1) + \frac{1}{2} \mathrm{LogGamma}(\thcomp{2} + 10|0, 1),\\
  \text{and, if}\,& d>2,\\
  \mathcal{L}(\thi) &=\begin{cases}
    \mathrm{LogGamma}(\thi|1, 1) \quad \text{for} \,\, 3 \le i \le \frac{d+2}{2},\\
    \mathrm{Normal}(\thi|0, 1) \qquad\,\,\,\, \text{for} \,\, \frac{d+2}{2} \le i \le d.
  \end{cases}
  \label{equ:loggamma_mix}
\end{split}
\end{equation}
Here the number of dimensions $d$ is even and the LogGamma distribution is
\begin{equation}
	\mathrm{LogGamma}(x|\alpha,\beta) = \frac{\e^{\beta x} \e^{-\e^x / \alpha}}{\alpha^\beta \Gamma(\beta)},
\end{equation}
where $\Gamma$ denotes the gamma function.

Our numerical tests all use uniform priors $\in [-30, 30]$ for each parameter.
As~\eqref{equ:loggamma_mix} and~\eqref{equ:gaussian} are both normalised to 1 and there is negligible posterior mass outside the prior, in both cases the evidence is almost exactly equal to the normalisation constant on the uniform prior --- i.e.
\begin{equation}
    \mathcal{Z}_\mathrm{true} = {60}^{-d}.\label{equ:z_true}
\end{equation}

\section{Diagnostic plots}\label{sec:plots}

\begin{figure}
\centering
\includegraphics[width=0.9\linewidth]{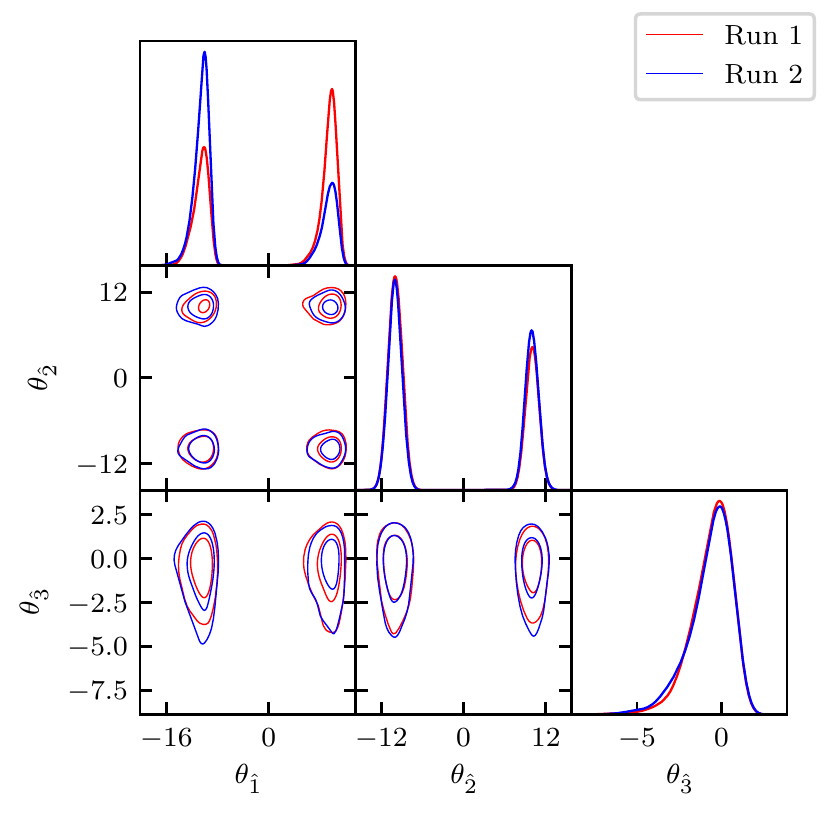}
\caption{Triangle plot of the posterior distributions for two nested sampling runs (red and blue lines), calculated using the 10-dimensional LogGamma mixture likelihood~\eqref{equ:loggamma_mix} and a uniform prior.
The on-diagonal plots show 1-dimensional marginalised posterior distributions on the first three parameters, and the remaining plots show calculated 2-dimensional 68\% and 95\% credible intervals on the joint posterior distribution.
The results for the two runs differ due to errors from both the intrinsic stochasticity of the nested sampling algorithm and implementation-specific effects.
Each nested sampling run has 250 live points, and uses the \PolyChord{} setting $\numrepeats{}=20$ --- this low setting is deliberately chosen to illustrate large implementation-specific effects.
}\label{fig:triangle}
\end{figure}

Before discussing quantitative diagnostics in~\Cref{sec:implementation_errors,sec:diagnostics}, we first introduce some diagnostic plots which illustrate nested sampling and its associated errors.
It is good practice for users of sampling software to represent their results visually, in order to assess if they are reasonable given background knowledge about the problem.
Many software packages exist for plotting 1- and 2-dimensional marginalised distributions from weighted samples using kernel density estimation.
As an example, \Cref{fig:triangle} shows posterior distributions for the LogGamma mixture likelihood~\eqref{equ:loggamma_mix}; this was made using \texttt{getdist} \citep{Lewis2015} with a zero-centred Gaussian kernel and the default settings.

While plots like \Cref{fig:triangle} are useful, it is unclear to what extent the differences between the two nested sampling runs are due to implementation-specific effects or merely what is expected from the stochasticity of the nested sampling algorithm.
Furthermore, these plots do not illustrate the distinctive manner in which nested sampling iterates towards higher likelihoods.
We therefore propose two additional diagnostic plots in \Cref{sec:bs_param_dist,sec:param_logx_diagram}, which can be calculated from nested sampling runs to show this extra information.
These are focused on distributions of parameters and so do not directly assess evidence calculations, but any significant inconsistencies in sample allocations observed between runs may also impact evidence estimates.

\subsection{Plotting the uncertainty on posterior distributions}\label{sec:bs_param_dist}

\begin{figure}
\centering
\begin{subfigure}{\linewidth}
    \centering
    \includegraphics{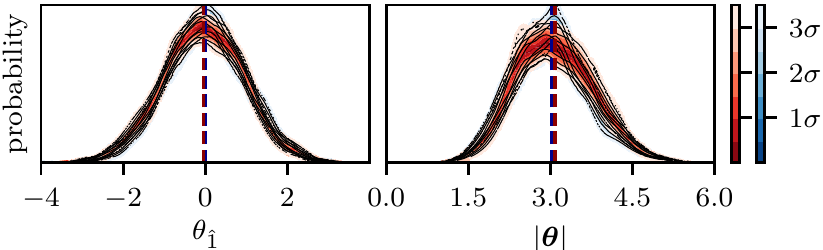}
    \subcaption{Posterior distributions of the first parameter and the radial coordinate $|\btheta|$ for a 10-dimensional Gaussian likelihood~\eqref{equ:gaussian}.}%
    \label{sub:bs_param_dist_gaussian}
\end{subfigure}
\begin{subfigure}{\linewidth}
    \centering
    \includegraphics{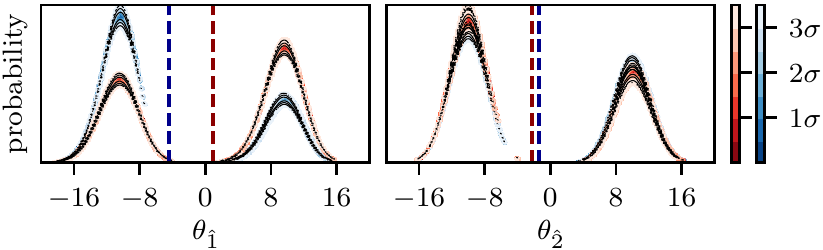}
    \includegraphics{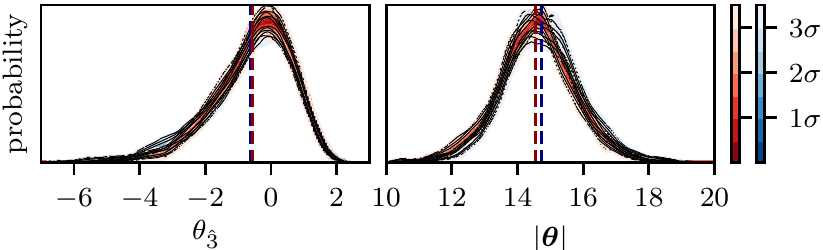}
    \subcaption{Posterior distributions of the first 3 parameters and $|\btheta|$ for a 10-dimensional LogGamma mixture likelihood~\eqref{equ:loggamma_mix}.
    The nested sampling runs are the same ones used in~\Cref{fig:triangle} with the corresponding colours.}%
    \label{sub:bs_param_dist_loggamma_mix}
\end{subfigure}
\caption{Diagrams of posterior distributions for two nested sampling runs (red and blue), showing the uncertainty due to the stochasticity of the nested sampling algorithm.
Each run uses 250 live points, and has $\numrepeats{}=20$ deliberately set to a low value to illustrate implementation-specific effects.
The coloured contours show iso-probability credible intervals on the marginalised posterior probability density function at each parameter value.
The dashed dark blue and dark red lines show the estimated posterior means of each parameter for the blue and red runs respectively.}%
\label{fig:bs_param_dist}
\end{figure}

The uncertainty on the posterior distributions due to nested sampling stochasticity can be estimated from a run by creating bootstrap resamples of the run using the procedure described in \citet[][Section 4]{Higson2017a}.
This uncertainty can be visually represented by plotting the distribution of the posteriors obtained from each resample (which is a nested sampling run) to give an {\em uncertainty distribution on the posterior distribution}.
Such plots can be used for assessing if the calculation error is sufficiently small for the given use case, and are illustrated in \Cref{fig:bs_param_dist}.
If they are of interest, the posterior distributions of functions of parameters can also be plotted; \Cref{sub:bs_param_dist_gaussian,sub:bs_param_dist_loggamma_mix} both show the radial coordinate $|\btheta| = {(\sum_i \thi^2)}^{1/2}$.
The coloured contours are plotted using the \texttt{fgivenx} package \citep{Handley2018fgivenx}.%
\footnote{When calculating plots like those in \Cref{fig:bs_param_dist}, the posterior distribution for each bootstrap replication must be calculated from the weighted samples without reducing them to evenly weighted samples in a stochastic manner --- such as by including each sample with probability proportional to its weight --- as this adds extra variation.
\nestcheck{} contains an implementation of 1-dimensional kernel density estimation which takes sample weights as an argument, and does not require conversion to evenly weighted samples.}

Plotting results from multiple runs on the same axis allows visual assessment of whether implementation-specific effects are present.
If posterior distributions differ by more than would be expected from their bootstrap sampling error distribution, then implementation-specific effects are likely to be the cause.
For example the top left panel of \Cref{sub:bs_param_dist_loggamma_mix}, in which the coloured distributions are clearly separated, suggests large implementation-specific effects are present in this case with the settings used.
\Cref{fig:bs_param_dist} deliberately uses low values for the \PolyChord{} \numrepeats{} and number of live points settings to illustrate implementation-specific effects; these effect can be reduced with a more appropriate choice of settings (discussed in~\Cref{sec:practical}).

\subsection{Plotting distributions of samples in $\log X$}\label{sec:param_logx_diagram}

We now propose a diagram to illustrate the distinctive manner in which a nested sampling run progresses by sampling from the prior with successively higher likelihood constraints, based on the discussion in \citet[][Section 3.1]{Higson2017a}.
This involves plotting sample parameters and weights against the fraction of the prior volume remaining, $X$, which is defined in~\eqref{equ:X_definition}.
A log scale is used as the shrinkage in $X$ at each step is exponential.

\begin{figure}
    \centering
    \includegraphics{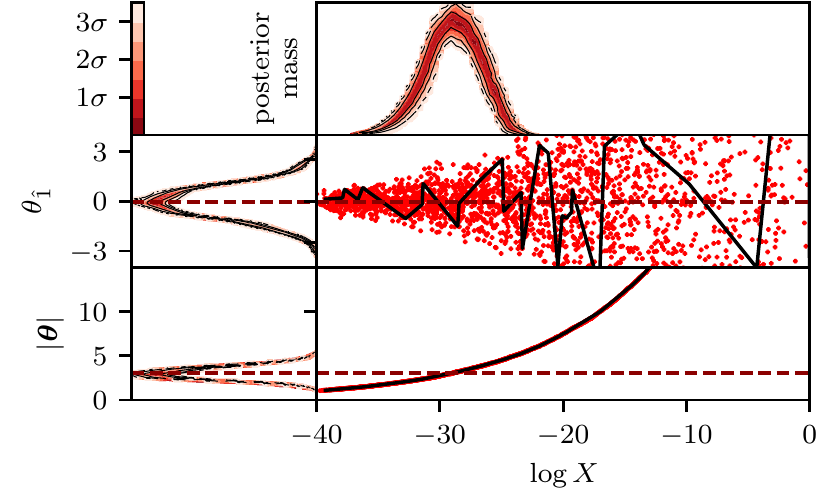}
    \caption{Diagram of samples' distributions in $\log X$ for a single run with a 10-dimensional Gaussian likelihood~\eqref{equ:gaussian}.
The top right panel shows the relative posterior mass (total weight assigned to all samples in that region) as a function of $\log X$.
The next two rows show the first parameter and the radial coordinate $|\btheta|$; for each the right panel plots its sampled values against $\log X$ and the left panel shows its posterior distribution in the same way as \Cref{sub:bs_param_dist_gaussian,sub:bs_param_dist_loggamma_mix}.
The coloured contours show iso-probability credible intervals on the marginalised posterior probability density function at each parameter or $\log X$ value.
The nested sampling run shown uses 250 live points and $\numrepeats{}=20$.
The solid black line shows the evolution of an individual thread (chosen at random).
The estimated mean value of the posterior distribution for each row is marked with a dashed line.}\label{sub:param_logx_gaussian}
\end{figure}

\begin{figure}
    \centering
    \includegraphics{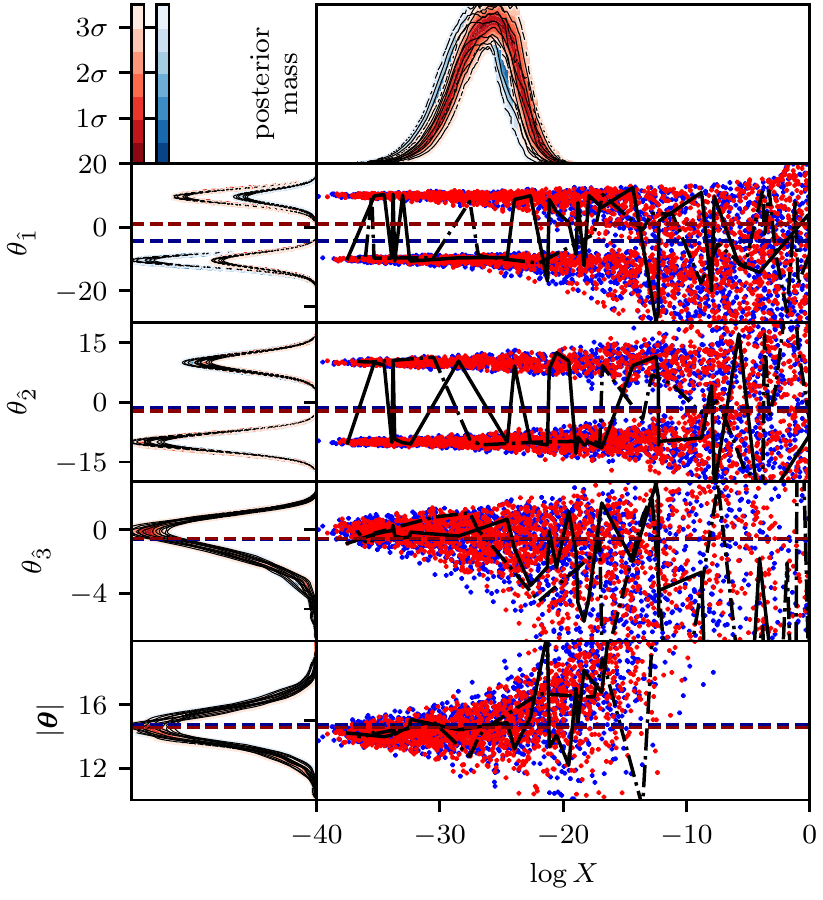}
    \caption{Diagram of samples' distributions in $\log X$ for two nested sampling runs from a 10-dimensional LogGamma mixture likelihood~\eqref{equ:loggamma_mix}.
The two runs (shown in red and blue) are the same ones used for \Cref{fig:triangle} and \Cref{sub:bs_param_dist_loggamma_mix}; each uses 250 live points and $\numrepeats{}=20$.
The top right panel shows the relative posterior mass (total weight assigned to all samples in that region) as a function of $\log X$.
The next four rows show the first 3 parameters and the radial co-ordinate $|\btheta|$; for each the right panel plots its sampled values against $\log X$ and the left panel shows its posterior distribution in the same way as \Cref{sub:bs_param_dist_gaussian,sub:bs_param_dist_loggamma_mix}.
The coloured contours show iso-probability credible intervals on the marginalised posterior probability density function at each parameter or $\log X$ value.
In each row, the estimated posterior means for the blue and red runs are shown with dashed dark blue and dark red lines.
The solid and dot dash black lines show the evolution of an individual thread chosen at random from the red and blue runs respectively.}%
    \label{sub:param_logx_loggamma_mix}
\end{figure}

In each plot the top right panel shows the relative posterior mass $\mathcal{L}(X)X$ (i.e.\ the weight assigned to samples in that $\log X$ region) on a relative scale; this is similar to \Cref{fig:ns_evidence}.
The $\log X$ co-ordinates of the samples are estimated statistically, with their uncertainty distribution displayed using coloured contours.
Each subsequent row represents a parameter or function of parameters, with the right panel showing the parameter value of each sample on the same $\log X$ scale.%
\footnote{The scatter plots in the right column of \Cref{sub:param_logx_gaussian,sub:param_logx_loggamma_mix} can be replaced with a colour plot of the estimated distribution of values at each $\log X$ using kernel density estimation \citep[similar to the colour distributions shown in Figure 3 of][]{Higson2017a}.
However doing this accurately is computationally challenging and requires a lot of samples, so simple scatter plots are typically more convenient for checking calculation results.}
The left panel is the same as the plots in the previous section (\Cref{sub:bs_param_dist_gaussian,sub:bs_param_dist_loggamma_mix}), and shows the posterior distribution on the parameter values on a shared scale with the left plot (including the uncertainty due to the stochasticity of the nested sampling algorithm).

Our proposed diagram is illustrated in \Cref{sub:param_logx_gaussian,sub:param_logx_loggamma_mix}.
The lower limit of the $\log X$ axis is chosen to include all points with non-negligible posterior mass, and the upper limit is set to 0 (the start of the nested sampling run).
The $y$-axis limits of the plots in the right column are simply chosen to include all samples with non-negligible posterior weight, or which are otherwise of interest.


In addition, the evolution of individual threads can be traced by drawing lines linking their constituent points.%
\footnote{Plots which trace individual threads in $\log X$ are also produced by the \dynesty{} dynamic nested sampling package. See \dynestyurl{} for more information.}
This shares similarities with MCMC trace plots but, unlike for a converged MCMC chain, the distribution of parameters changes as the algorithm iterates over different $\log X$ values.
Furthermore, as the algorithm progresses towards lower values of $\log X$ it moves from right to left in the diagram; in MCMC trace plots, chains typically move from left to right.

\Cref{sub:param_logx_gaussian,sub:param_logx_loggamma_mix} are useful for visualising the nested sampling process and parts of the posterior such as degeneracies and modes with which nested sampling software may struggle.
Furthermore if additional information about the posteriors is available, such as that they should have certain symmetries or be unimodal, this type of diagram can be useful in working out where the sampler is not behaving as expected.
For example~\Cref{sub:param_logx_loggamma_mix} clearly shows the multi-modality of the LogGamma mixture likelihood, as well as giving an indication of when in the nested sampling process the modes separate.
In addition the bottom right panel of~\Cref{sub:param_logx_gaussian} shows that the radial coordinate $|\btheta|$ has negligible spread at any given $\log X$ value in this case; this is due to the likelihood and prior's spherical symmetry.

Furthermore, multiple nested sampling runs can be added to the same axis --- as shown in \Cref{sub:param_logx_loggamma_mix}.
This allows comparison of where runs differ; for example one may be able to see on the plot that one of the runs had missed a mode which the other run found (although in \Cref{sub:param_logx_loggamma_mix} the samples from the two runs overlap).
One can also see from \Cref{sub:param_logx_loggamma_mix} that the two runs agree closely on the relative weights assigned at different $\log X$ values (top panel), meaning that the difference between the posterior distributions (left panels) is due to the parameter values sampled in each $\log X$ region rather than the distribution of posterior mass.%
\footnote{It is common for the parameter values sampled to be the main difference between parameter estimation calculations using different runs, as only the relative weights of points affect the calculation \citep[see][for more details]{Higson2017a}.}

\section{Estimating implementation-specific effects}\label{sec:implementation_errors}

Following the diagnostics plots of the previous section, the remainder of this paper discusses quantitatively measuring implementation-specific effects.
The total error on nested sampling calculations can be estimated by measuring the variation of results when a calculation is repeated multiple times, as this includes both implementation-specific effects and the intrinsic stochasticity of the algorithm.
This provides a lower bound on the total error, but will underestimate it in the case that implementation-specific effects cause calculation results to be systematically biased.

While the nature of implementation-specific effects depends on the specific software used, they are very likely to be uncorrelated with the errors from the stochasticity of the nested sampling algorithm --- which can be calculated using the bootstrap resampling approach.
Assuming that they are indeed uncorrelated, the variance in posterior inferences (such as the calculated values of parameter means or the Bayesian evidence) due to implementation-specific effects $\sigma_\mathrm{imp}^2$ is related to the variance estimated from bootstrap resampling $\sigma_\mathrm{bs}^2$ and the sample variance of calculation results $\sigma_\mathrm{values}^2$ by the standard relation for the sum of the variances of uncorrelated random variables (the Bienaym\'e formula)
\begin{equation}    
    \sigma^2_\mathrm{values}
    =
    \sigma^2_\mathrm{bs}
    +
    \sigma^2_\mathrm{imp}.
\end{equation}
Using this result, we propose calculating the standard deviation of the uncertainty distribution due to implementation-specific effects $\sigma_\mathrm{imp}$ as
\begin{equation}    
    \sigma_\mathrm{imp} =
\begin{cases}
    \sqrt{\sigma^2_\mathrm{values} - \sigma^2_\mathrm{bs}}
    &\text{if}\, \sigma^2_\mathrm{values} > \sigma^2_\mathrm{bs},
    \\
    0
    &\text{otherwise}.
    \label{equ:implementation_error}
\end{cases}
\end{equation}
To summarise: here $\sigma_\mathrm{values}$ is the observed sample standard deviation of results, $\sigma_\mathrm{bs}$ represents the standard deviation we would expect if the nested sampling algorithm was performed perfectly, and $\sigma_\mathrm{imp}$ represents the implementation-specific effects causing the difference.

If a number of nested sampling runs are available, the implementation-specific effects on calculations of scalar quantities such as the mean and median of parameters can be calculated directly from~\eqref{equ:implementation_error} and compared to the variation of results.
One can also estimate the fraction of the observed variation which is due to implementation-specific effects $\sigma_\mathrm{imp} / \sigma_\mathrm{values}$ --- when implementation-specific effects are large this is easy to measure accurately as the variation of results is much greater than the bootstrap error estimates and
\begin{equation}    
    \frac{\sigma_\mathrm{imp}}{\sigma_\mathrm{values}}
    =
    \frac{\sqrt{\sigma^2_\mathrm{values} - \sigma^2_\mathrm{bs}}}{\sigma_\mathrm{values}}
    =
    1 - \frac{\sigma_\mathrm{bs}}{2 \sigma_\mathrm{values}} + \mathcal{O}\left(\frac{\sigma_\mathrm{bs}^2}{\sigma_\mathrm{values}^2}\right).
    \label{equ:implementation_error_frac}
\end{equation}
The number of runs required to estimate $\sigma_\mathrm{imp}$ is primarily determined by the accuracy of the sample standard deviation $\sigma_\mathrm{values}$.
\citet{Ahn2003} give a formula for the fractional uncertainty of the sample standard deviation as a function of the number of data points; for computationally expensive problems in our research, we typically use $\sim10$ runs to estimate $\sigma_\mathrm{imp}$.
In practice $\sigma_\mathrm{bs}$ makes a negligible contribution to the uncertainty on $\sigma_\mathrm{imp}$; it can be estimated accurately from a single run, and the accuracy can be further improved by averaging estimates from all the runs available.

\begin{figure}
\centering
\includegraphics{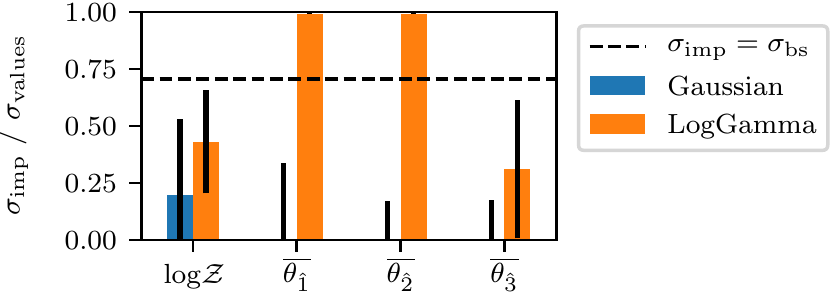}
\caption{Ratios of estimated implementation-specific effects~\eqref{equ:implementation_error} to variation of results for 10-dimensional Gaussian~\eqref{equ:gaussian} and LogGamma mixture~\eqref{equ:loggamma_mix} likelihoods.
The dashed horizontal line at $\sigma_\mathrm{imp}/\sigma_\mathrm{values}=\frac{1}{\sqrt{2}}$ shows the level where implementation-specific effects and the stochasticity of the nested sampling algorithm make equal contributions to the total error; ratios above this value imply the majority of the error is due to implementation-specific effects.
Each bar is calculated using 100 \PolyChord{} runs, each with 250 live points and $\numrepeats{}=50$.
Results are shown for the log-evidence, the mean of the two parameters, the mean radial coordinate and the second moment of $\thcomp{1}$.
The numerical results plotted in this figure are given in \Cref{tab:gaussian,tab:loggamma_mix} in Appendix~\ref{app:tables}.}%
\label{fig:imp_error_bars}
\end{figure}

\Cref{fig:imp_error_bars} shows the ratio of the inferred implementation error to the total variation of results for 100 nested sampling runs using 10-dimensional Gaussian~\eqref{equ:gaussian} and LogGaussian mixture~\eqref{equ:loggamma_mix} likelihoods.
As for \Cref{fig:triangle,fig:bs_param_dist,sub:param_logx_gaussian,sub:param_logx_loggamma_mix} we use the \PolyChord{} setting $\numrepeats{}=20$, which is deliberately chosen to be low in order to illustrate implementation-specific effects.
The numerical results plotted in \Cref{fig:imp_error_bars} are given in \Cref{tab:gaussian,tab:loggamma_mix} in Appendix~\ref{app:tables}, along with the absolute values of the variation of results, root-mean-squared-errors and implementation error estimates.
With these \PolyChord{} settings, implementation-specific effects are the dominate source of parameter estimation errors for the LogGamma mixture likelihood.
However, the implementation fraction of the error for the log-evidence calculations is significantly lower than for parameter estimation; this is because errors from the stochasticity of the nested sampling algorithm are much larger for evidence calculation than for parameter estimation.

The mean calculated value of $\log \mathcal{Z}$ for the LogGamma mixture likelihood~\eqref{equ:loggamma_mix}, shown in~\Cref{tab:loggamma_mix}, differs by $0.10\pm0.03$ from the true value from~\eqref{equ:z_true} of $\log \Z_\mathrm{true} = -d \log(60)$.
This systematic bias is due to \PolyChord{} failing to consistently explore the posterior in this challenging case with the deliberately low setting $\numrepeats$ setting used --- it can be reduced by increasing \numrepeats.
However despite the bias, our approach successfully detected implementation-specific effects in this case.
Furthermore, using the true value, we can calculate implementation-specific effects by using the root-mean-squared-error (RMSE) in~\eqref{equ:implementation_error}:
\begin{equation}    
    \sigma_\mathrm{imp,RMSE} =
\begin{cases}
    \sqrt{\mathrm{RMSE}^2 - \sigma^2_\mathrm{bs}}
    &\text{if}\, \mathrm{RMSE}^2 > \sigma^2_\mathrm{bs},
    \\
    0
    &\text{otherwise}.
    \label{equ:implementation_error_rmse}
\end{cases}
\end{equation}
In this case the estimated $\sigma_\mathrm{imp} / \sigma_\mathrm{values}$ ratio of $0.43\pm0.23$ shown in \Cref{fig:imp_error_bars} is only a small underestimate compared to $\sigma_\mathrm{imp,RMSE} / \mathrm{RMSE} = 0.50 \pm 0.14$.
Assessing results for systematic bias when the true value of the quantity is not available is discussed in \Cref{sec:bias}.

\citet{Skilling2006} recommends that inferences from multiple nested sampling runs are made by combining them into a single run rather than simply averaging the results from each run, as this allows more accurate estimation of sample weights.
If implementation-specific effects are negligible then uncertainty estimates can be calculated from the combined run using standard techniques, but this will be inaccurate if implementation-specific effects are the dominant source of error.
In the latter case, the approximate error on the combined inference $\sigma_\mathrm{combined}$ from $N$ nested sampling runs with the same settings can be roughly estimated as
\begin{equation}
    \sigma_\mathrm{combined} = \sigma_\mathrm{values} / \sqrt{N}.
    \label{equ:unc_comb}
\end{equation}
This may be an overestimate as it does not including the benefits of combining the runs, but in practice this effect is likely to be small compared to the uncertainty in the sample standard deviation of the separate runs $\sigma_\mathrm{values}$ unless $N$ is very large.

\section{Diagnostic tests for when few runs are available}\label{sec:diagnostics}

For computationally expensive problems there may not be enough nested sampling runs available to calculate the implementation-specific effects directly using the method described in the previous section.
In \Cref{sec:thread_diagnostics,sec:bootstrap_diagnostics} we therefore consider diagnostics which assess whether two nested sampling runs have consistently explored a parameter space while accounting for the stochastic nature of the nested sampling algorithm.
Due to the relatively small amount of information available in this case, it is useful to also consider qualitative comparisons using diagnostic plots of the types shown in \Cref{sec:plots} as well as any problem-specific knowledge of what the results should be.
If $N > 2$ runs are available then $\binom{N}{2}$ pairwise tests can be computed and their results combined for greater accuracy.

\subsection{Testing for correlations between threads}\label{sec:thread_diagnostics}

We now introduce a test to assess whether nested sampling software is consistently exploring a posterior by comparing the statistical properties of the set of constituent threads (single live point runs) of two nested sampling runs.
Each thread represents a valid nested sampling run and can be used to make posterior inferences about quantities such as the evidence and the mean and median of parameters.
The actual values calculated from each thread will have large errors due their small number of samples, but this does not matter for testing if the distributions of values obtained from each run's threads are consistent.

We propose applying the 2-sample Kolmogorov-Smirnov (KS) test \citep{Massey1951} to different runs' constituent threads by using each thread to calculate an estimate of a scalar quantity of interest (such as parameter means or the Bayesian evidence $\mathcal{Z}$) with the following procedure:
\begin{enumerate}
    \item divide the first nested sampling run into its $n_1$ constituent threads, and calculate an estimate of the quantity from each;
    \item divide the second nested sampling run into its $n_2$ constituent threads, and calculate an estimate of the quantity from each;
    \item apply the 2-sample KS test to the $n_1$ and $n_2$ values calculated from the first and second runs respectively.
\end{enumerate}
As a test statistic for distributions $p(x)$ and $q(x)$, the KS test uses the maximum distance between their cumulative distributions $F_p(x)$ and $F_q(x)$
\begin{equation}\label{equ:ks_statistic}
    D_{p,q} = \sup_x |F_p(x) - F_q(x)|,
\end{equation}
where $\sup$ is the supremum.
If $n_1$ and $n_2$ samples from $p(x)$ and $q(x)$ respectively are used, the corresponding $p$-values are
\begin{equation}
    \alpha = 2 \exp \left( - \frac{2  n_1 n_2}{n_1 + n_2} D_{p,q}^2\right).
\end{equation}
In this case the $p$-value produced represents the probability of observing a KS statistic $D_{p,q}$ of this size or greater if the threads in the two runs were drawn from the same distribution.
A $p$-value close to zero implies that the values obtained from the threads in the two runs are statistically inconsistent, and hence that implementation-specific effects are likely to be present.
This procedure can also be used with other distribution-free tests such as the 2-sample Anderson-Darling test \citep{Scholz1987} as an alternative to the KS test.

\begin{figure}
\centering
\begin{subfigure}{\linewidth}
    \centering
    \includegraphics{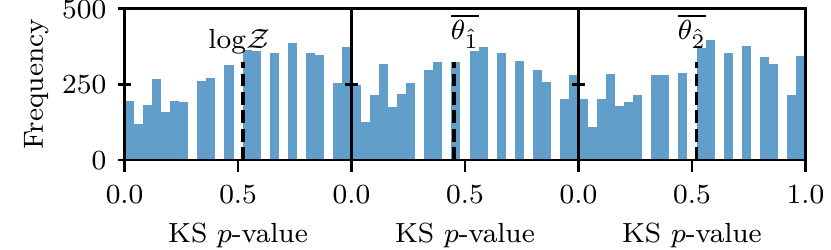}
    \subcaption{10-dimensional Gaussian likelihood~\eqref{equ:gaussian}.}%
    \label{sub:hist_pvalue_gaussian}
\end{subfigure}
\begin{subfigure}{\linewidth}
    \centering
    \includegraphics{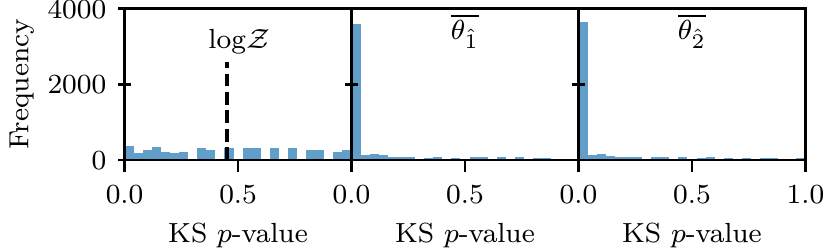}
    \subcaption{10-dimensional LogGamma mixture likelihood~\eqref{equ:loggamma_mix}.}%
    \label{sub:hist_pvalue_loggamma_mix}
\end{subfigure}
\caption{Distributions of KS $p$-values from pairwise comparison of different runs' constituent threads, using $\log \mathcal{Z}$ and the first two parameters.
A $p$-value of 0 means the quantities calculated from threads in the two runs are from different distributions, implying the threads within each run are correlated with each other and implementation-specific effects are present.
The black dashed line shows the median $p$-value for each plot.
The nested sampling runs are the same ones that were used for \Cref{fig:imp_error_bars} --- the 100 runs allow $\binom{100}{2} = 4,950$ pairwise statistics to be computed.}%
\label{fig:hist_pvalue}
\end{figure}

\Cref{fig:hist_pvalue} shows distributions of the $p$-values computed by applying this procedure to different pairs of nested sampling runs.
For the LogGamma mixture likelihood the median $p$-values for $\thmean{1}$ and $\thmean{2}$ are $2\times10^{-4}$ and $5\times10^{-5}$ respectively, strongly suggesting that implementation-specific effects are present (in agreement with~\Cref{fig:imp_error_bars}).
However, the approach is not able to detect significant evidence of implementation-specific effects in $\log \mathcal{Z}$ calculations, as implementation-specific effects comprise only a fraction of the total variation of results in this case so the pairs of runs do not provide enough information.

In addition there are many quantities which can be tested --- for example the Bayesian evidence and the mean, median, higher moments and credible intervals of each parameter.\footnote{Tests on functions of the same parameter will not be independent.}
Considering a number of quantities allows sensitive testing for implementation-specific errors from only two runs, even if the implementation-specific effects are smaller than in the LogGamma mixture case.
One could also test multiple quantities together using a multi-dimensional KS test, although this is challenging as there is no unique order for quantity values in more than 1 dimension --- see \citet{Fasano1987} for a more detailed discussion.
An alternative is to use multiple hypothesis testing with $p$-value corrections, for example with the Holm-Bonferroni method \citep{Holm1979}.

For \MultiNest{} runs using the setting \texttt{mmodal=True}, when a new mode is recognised, the run is split and live points assigned to the mode remain in that mode and evolve independently from the remainder of the run.
As a result, even when there are no implementation-specific effects, the threads within such a run are not independently drawn from the same distribution and the KS test will not give correct $p$-values.
The test is valid for \PolyChord{} runs and \MultiNest{} runs with \texttt{mmodal=False} as in these cases threads move between modes; this can be seen in~\Cref{sub:param_logx_loggamma_mix}.

It is important to note that the KS $p$-value only determines whether implementation-specific effects are present and does not provide information about the size of implementation error, which must be assessed to determine if they are problematic for a given use case.%
\footnote{In particular with enough data (threads) one can get very low $p$-values even if the implementation-specific effects are relatively small and/or not important for the practical problem being examined.}
This can be done with the help of bootstrap resamples, as discussed in the next section.

\subsection{Testing the consistency of sampling error distributions}\label{sec:bootstrap_diagnostics}

\begin{figure}
\centering
\begin{subfigure}{\linewidth}
    \centering
    \includegraphics{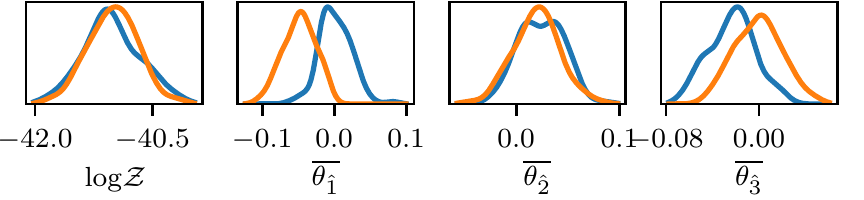}
    \subcaption{10-dimensional Gaussian likelihood~\eqref{equ:gaussian}.}%
    \label{sub:1dkde_gaussian}
\end{subfigure}
\begin{subfigure}{\linewidth}
    \centering
    \includegraphics{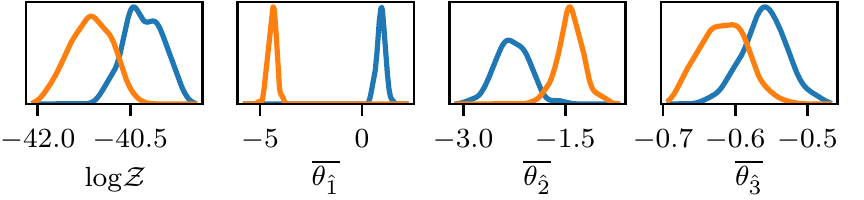}
    \subcaption{10-dimensional LogGamma mixture likelihood~\eqref{equ:loggamma_mix}.}%
    \label{sub:1dkde_loggamma_mix}
\end{subfigure}
\caption{Plots of the sampling errors distribution calculated from bootstrap resampling threads for different quantities.
Each plot shows 2 nested sampling runs (represented by different line colours), each with 250 live points and $\numrepeats{}=20$.
The kernel density estimation of the posterior distributions use a Gaussian kernel with the bandwidth selected using Scott's rule \citep{Scott2015}.
These plots are designed for use when the true values are not available (although in this case the true values for the distributions shown can be found in \Cref{tab:gaussian,tab:loggamma_mix}).}%
\label{fig:1dkde}
\end{figure}

Our second diagnostic assesses whether calculations of scalar quantities from the two different runs differ by more than would be expected given the estimated uncertainties from the intrinsic stochasticity of the nested sampling algorithm.
These uncertainty distributions on posterior point estimates can be calculated from bootstrap resamples using the method described in~\citet{Higson2017a}, and are illustrated in \Cref{sub:1dkde_gaussian,sub:1dkde_loggamma_mix}.
This has some similarities with \Cref{sub:bs_param_dist_gaussian,sub:bs_param_dist_loggamma_mix} but considers only errors on single numbers (such as the means of parameters shown by dashed vertical lines in those figures) rather than on whole posterior distributions.
As a result this approach can also be applied to the Bayesian evidence $\mathcal{Z}$, which is a number rather than a distribution.

Bootstrapped point estimates can be qualitatively compared across runs using plots like \Cref{fig:1dkde}, or the statistical distance between the distributions can be quantified.
As with the comparisons of threads in \Cref{sec:thread_diagnostics} it may be hard to draw conclusions from any one quantity, but the two runs can be compared using many different posterior estimates.
Quantification may be more convenient than plotting graphs when comparing many different quantities or pairs of runs.

\begin{figure}
\centering
\begin{subfigure}{\linewidth}
    \centering
    \includegraphics{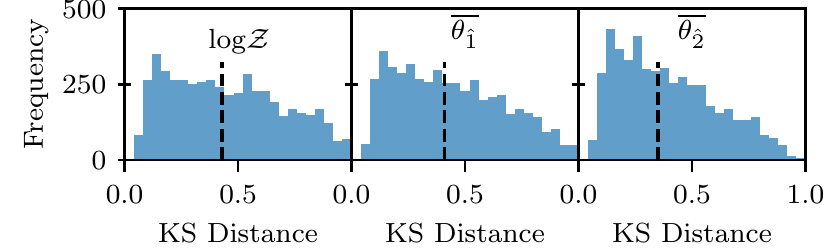}
    \subcaption{10-dimensional Gaussian likelihood~\eqref{equ:gaussian}.}%
    \label{sub:hist_ks_dist_gaussian}
\end{subfigure}
\begin{subfigure}{\linewidth}
    \centering
    \includegraphics{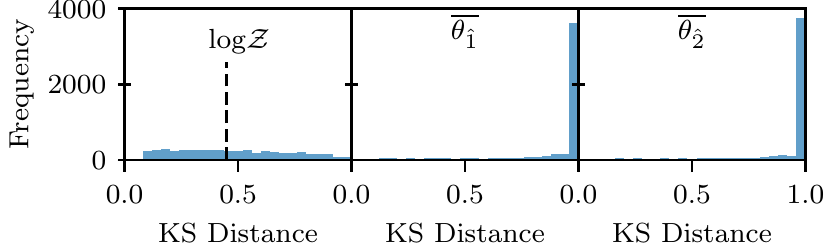}
    \subcaption{10-dimensional LogGamma mixture likelihood~\eqref{equ:loggamma_mix}.}%
    \label{sub:hist_ks_dist_loggamma_mix}
\end{subfigure}
\caption{Distributions of KS statistical distances~\eqref{equ:ks_statistic} between bootstrap uncertainty distributions on point estimates the type shown in~\Cref{fig:1dkde}.
For each likelihood, the 3 columns show results for $\log \mathcal{Z}$ calculations and for the mean of the parameters $\thcomp{1}$ and $\thcomp{2}$.
The nested sampling runs are the same ones that were used for \Cref{fig:imp_error_bars}; the 100 runs are compared pairwise to give $\binom{100}{2} = 4,\!950$ KS statistical distances for each quantity.
A KS statistic of close to 1 means there is little overlap between the distributions, implying that the differences in the runs' values cannot be explained by the intrinsic stochasticity of the nested sampling algorithm and that implementation-specific effects are present.
The black dashed line shows the median KS distance for each plot.
}%
\label{fig:hist_ks_dist}
\end{figure}

We use the KS statistic~\eqref{equ:ks_statistic} as a statistical distance measure; this constitutes a metric as it is non-negative, zero if and only if the distributions are equal, symmetric and satisfies the triangle inequality.
Its numerical values are also easy to interpret, with a value of 0 meaning the distributions are the same and a value of 1 meaning they do not overlap.
KS statistical distances between bootstrapped posterior point estimates from different pairs of nested sampling runs are shown in \Cref{fig:hist_ks_dist}.
These distributions show strong evidence for implementation-specific effects in parameter estimation for the LogGamma mixture case, with calculations of $\thmean{1}$ and $\thmean{2}$ having $65.7\%$ and $67.9\%$ of their pairwise statistical distances equalling 1 respectively.
These estimates are particularly sensitive to changes in the relative weighting of different modes in the posterior.
However, as for the diagnostic introduced in \Cref{sec:thread_diagnostics}, two runs do not provide enough information to detect the relatively weaker implementation-specific effects in the LogGamma mixture $\log \mathcal{Z}$ estimates.

The KS statistical distances are more difficult to interpret than the $p$-values in~\Cref{sec:thread_diagnostics}, but have the advantage that together with plots like~\Cref{fig:1dkde} they contain information about the size of any implementation-specific effects.
In this context, the KS statistic values are simply used as a distance measure and cannot be interpreted as $p$-values.
This is because, even without implementation-specific effects, nested sampling runs will differ due to the stochasticity of the algorithm, and these differences mean bootstrap resamples of different runs are drawn from different distributions.

\section{Implementation-specific effects in practice}\label{sec:practical}

Having introduced our diagnostic tests, we now empirically test how different software settings and problem dimension affect the size of implementation-specific effects.
As an example we use \PolyChord{}, but we intend this section to be informative for users of other software packages such as \MultiNest{} and \dyPolyChord{}.
The section finishes with practical advice for software users.

\subsection{Effect of sampling efficiency settings}

\begin{figure}
\centering
\includegraphics{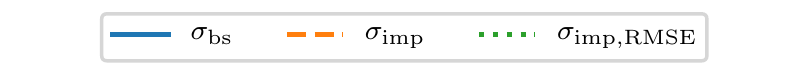}
\begin{subfigure}{0.49\linewidth}
    \centering
    \includegraphics{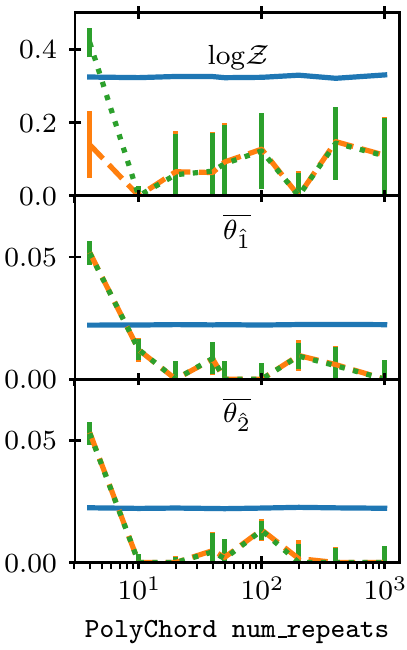}
    \subcaption{10-dimensional Gaussian likelihood~\eqref{equ:gaussian} with a uniform prior.}
\end{subfigure}\hfill
\begin{subfigure}{0.49\linewidth}
    \centering
    \includegraphics{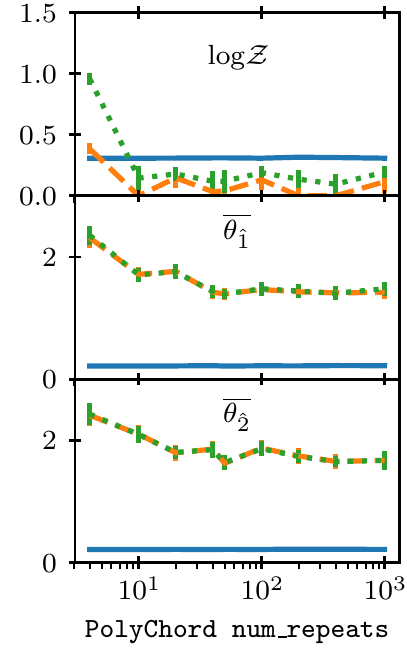}
    \subcaption{10-dimensional LogGamma mixture~\eqref{equ:loggamma_mix} with a uniform prior.}%
    \label{sub:line_nrepeats_loggamma_mix}
\end{subfigure}
\caption{The effect of \PolyChord{}'s \numrepeats{} setting on results errors; each subfigure shows calculations of the log-evidence and the mean of the first two parameters.
Results for every $\numrepeats$ value were calculated using 100 nested sampling runs, each with 250 live points.
Blue solid lines show the mean bootstrap error estimate and orange dashed lines show implementation-specific effect estimates from~\eqref{equ:implementation_error}.
Green dotted lines show the implementation-specific effects calculated using the root-mean-squared-error~\eqref{equ:implementation_error_rmse}; where the green dotted and orange dashed lines are equal, there is no systematic bias in the results.
Error bars show the uncertainty on results for each \numrepeats{} value considered.
}\label{fig:line_nrepeats}
\end{figure}

Nested sampling software packages typically have settings controlling the process of sampling within a hard likelihood constraint which can reduce implementation-specific effects at the cost of increased computation.
\PolyChord{} and \dyPolyChord{} both have a \numrepeats{} setting which controls the number of slice samples taken before sampling each new live point --- increasing this value reduces correlation between points and increases the accuracy with which they perform the nested sampling algorithm.
Other examples of similar parameters include \MultiNest's \efr{}, which controls the efficiency of its rejection sampling algorithm by determining the size of the ellipsoid within which \MultiNest{} samples.
If \efr{} is lowered, samples are drawn from a larger ellipsoid, increasing the rejection rate whilst consequently decreasing the chance of missing part of the parameter space within the iso-likelihood contour.
Hence, in contrast with \numrepeats, implementation-specific effects are made smaller by {\em reducing\/} \efr{}.

\Cref{fig:line_nrepeats} shows the effect on calculation errors of \PolyChord's \numrepeats{} setting.
As expected, we see that as \numrepeats{} is increased the implementation-specific effects are reduced --- showing \PolyChord{} is performing the nested sampling algorithm with increasing accuracy.
However, the $\numrepeats$ value required for implementation-specific effects to be a small fraction of the total error is highly problem dependent, even for the same number of dimensions.
For the 10-dimensional Gaussian likelihood $\numrepeats = 10$ is easily sufficient, but for the challenging 10-dimensional LogGamma likelihood $\numrepeats >10^3$ is needed.
\numrepeats{} can be tuned by, for example, doubling it until results show small implementation errors.
In principle a sufficiently high $\numrepeats$ value can make such errors negligible even for challenging likelihoods, but this will become impractically computationally expensive and gives diminishing returns in cases like the LogGamma mixture shown in \Cref{sub:line_nrepeats_loggamma_mix}.
Once \numrepeats{} is high enough that the calculations are not systematically biased, simply repeating the calculation many times is more efficient at improving accuracy.
One can check for such a bias by assessing if the mean value of results changes when \numrepeats{} is increased (if a bias is present, increasing \numrepeats{} should reduce it).

\subsection{Effect of the number of live points}

\begin{figure}
\centering
\includegraphics{img/line_plot_legend.pdf}
\begin{subfigure}{0.49\linewidth}
    \centering
    \includegraphics{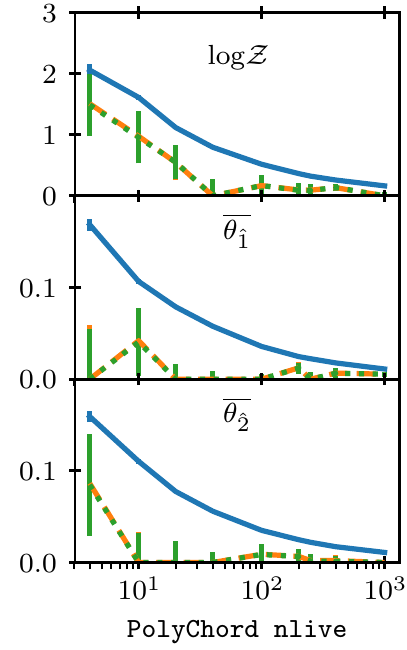}
    \subcaption{10-dimensional Gaussian likelihood~\eqref{equ:gaussian} with a uniform prior.}
\end{subfigure}\hfill
\begin{subfigure}{0.49\linewidth}
    \centering
    \includegraphics{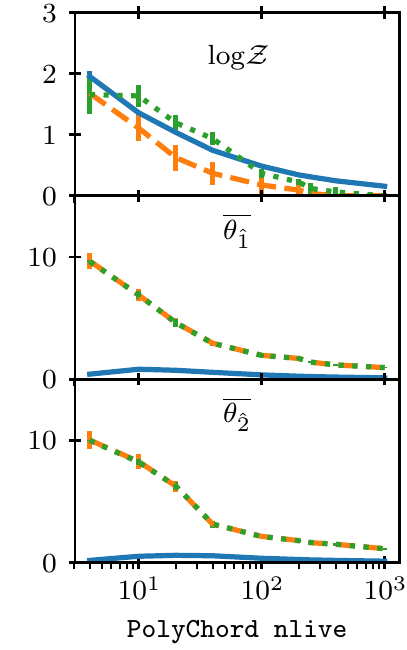}
    \subcaption{10-dimensional LogGamma mixture~\eqref{equ:loggamma_mix} with a uniform prior.}%
\end{subfigure}
\caption{The effect of the number of live points on errors in \PolyChord{} calculations; the two subfigures both show calculations of the log-evidence and the mean of the first two parameters.
Results for each number of live points considered were calculated using 100 nested sampling runs with $\numrepeats{}=10$.
Blue solid lines show the mean bootstrap error estimate and orange dashed lines show implementation-specific effect estimates from~\eqref{equ:implementation_error}.
Green dotted lines show the implementation-specific effects calculated using the root-mean-squared-error~\eqref{equ:implementation_error_rmse}; where the green dotted and orange dashed lines are equal, there is no systematic bias in the results.
Error bars show $1\sigma$ uncertainties on results for each number of live points considered.
}\label{fig:line_nlive}
\end{figure}
 
In addition to software specific settings, the main choice a nested sampling user must make is the number of live points, which controls the resolution of sampling and is proportional to the expected number of samples produced.
For simplicity we consider only runs with a constant number of live points $n$, although our conclusions also apply to dynamic nested sampling \citep{Higson2017b} --- in which the number of live points varies to increase calculation accuracy.
Furthermore, \nestcheck{} is compatible with the output of several dynamic nested sampling software packages including \dyPolyChord{}, \dynesty{}\footnote{See \dynestyurl{} for more information.} and \perfectns{}.

The changes in calculation errors with changes in the number of live points used is shown in \Cref{fig:line_nlive}.
As expected, increasing the number of live points reduces the implementation-specific effects, as well as the errors from the stochasticity of the nested sampling algorithm (measured by bootstrap resampling) which are approximately proportional to $1/\sqrt{n}$.
The fraction of the total error made up by implementation-specific effects does not necessarily decrease with increased $n$ --- this depends on how the implementation-specific effects scale with $n$.
For the Gaussian likelihood, implementation-specific effects cause only a small part of the total variation of results, whereas for the more challenging LogGamma mixture likelihood they are the main source of errors.

Given that increasing $n$ reduces both implementation-specific effects and errors from the stochasticity of the nested sampling algorithm, this is often a better way to reduce total errors for the same computational cost than increasing \numrepeats{}.
However it may not reduce the fraction of errors caused implementation-specific effects.
Consequently, techniques for estimating nested sampling errors which do not account for implementation-specific effects may still underestimate the total uncertainties.

\subsection{Calculation results with a systematic bias}\label{sec:bias}

\Cref{fig:line_nrepeats,fig:line_nlive} show that for $\log \mathcal{Z}$ calculations, if \nlive{} and \numrepeats{} are set too low, estimates of the implementation-specific effects using the standard deviation of results and the root-mean-square error can start to differ.
This is due to the algorithm failing to fully explore the posterior and iterating inwards too quickly, which leads to a systematic bias in $\log \mathcal{Z}$ \citep[this is discussed in detail in][]{Buchner2016a}.
The \nlive{} and \numrepeats{} settings required to remove the bias depend on the posterior, with challenging multimodal or degenerate posteriors needing more samples (as for implementation-specific effects).
The challenging LogGamma mixture likelihood shows a bias with the \PolyChord{} settings used (as shown in \Cref{tab:loggamma_mix} in Appendix~\ref{app:tables}), but this is small compared to the standard deviation of calculation results and can be reduced by increasing \numrepeats{} or the number of live points.
Systematic biases in a parameter estimation calculations are also possible with inappropriate settings, but in the authors' experience this is much rarer.

The failure to fully explore the posterior which causes a systematic bias typically also results in differences between runs which are not explained by the stochasticity of the nested sampling algorithm --- these implementation-specific effects can be detected the diagnostic tests presented in this paper.
However, the bias causes these diagnostics to underestimate the size of the implementation-specific effects.
If significant implementation-specific effects are detected in runs and the results of $\log \mathcal{Z}$ calculations are of interest, one can check for bias by repeating the calculation with higher \nlive{} and \numrepeats{} settings and checking if the mean calculated result changes.

\subsection{Effect of dimensionality}\label{sec:dimensions}

\begin{figure}
\centering
\includegraphics{img/line_plot_legend.pdf}
\begin{subfigure}{0.49\linewidth}
    \centering
    \includegraphics{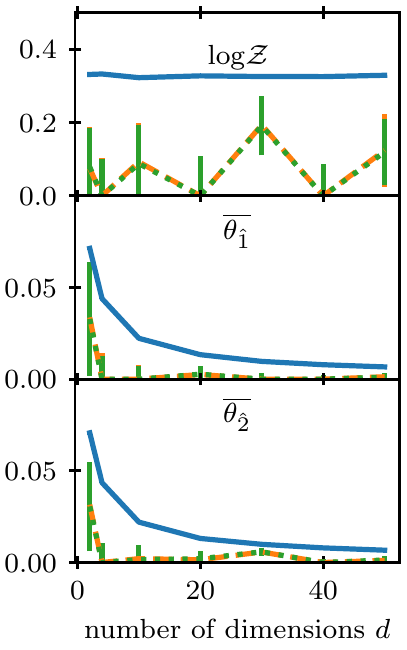}
    \subcaption{10-dimensional Gaussian likelihood~\eqref{equ:gaussian} with a uniform prior.}
\end{subfigure}\hfill
\begin{subfigure}{0.49\linewidth}
    \centering
    \includegraphics{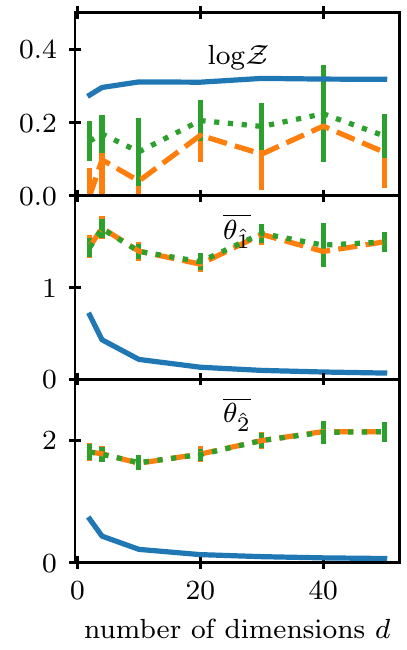}
    \subcaption{10-dimensional LogGamma mixture~\eqref{equ:loggamma_mix} with a uniform prior.}%
    \label{sub:line_ndim_loggamma_mix}
\end{subfigure}
\caption{The effect of increasing the dimension $d$ on errors in \PolyChord{} calculations: each subfigure shows calculations of the log-evidence and the mean of the first two parameters.
Results for every dimension $d$ use $25 \times d$ live points and the \PolyChord{} setting $\numrepeats=5\times d$.
Blue solid lines show the mean bootstrap error estimate and orange dashed lines show implementation-specific effect estimates from~\eqref{equ:implementation_error}.
Green dotted lines show the implementation-specific effects calculated using the root-mean-squared-error~\eqref{equ:implementation_error_rmse}; where the green dotted and orange dashed lines are equal, there is no systematic bias in the results.
Error bars show $1\sigma$ uncertainties on results for different numbers of dimensions.
}\label{fig:line_ndim}
\end{figure}

\Cref{fig:line_ndim} shows implementation errors for the Gaussian and LogGamma mixture likelihoods for different numbers of dimensions $d$.
Each calculation uses $25 \times d$ live points and $\numrepeats = 5 \times d$ (the default settings in \PolyChord's Python interface).
These are proportional to $d$ in order to give approximately constant errors in $\log Z$ \citep{Handley2015b}, with the additional samples produced for higher $d$ leading to lower parameter estimation errors.
With these settings, as $d$ increases, our plot shows no strong upwards or downwards trend in the implementation error.
Furthermore, the small bias in the $\log \mathcal{Z}$ calculation results for the LogGamma mixture likelihood (shown by the difference between the green dotted and orange dashed lines in the top panel of~\Cref{sub:line_ndim_loggamma_mix}) remains much smaller than the standard deviation of the results values $\sigma_\mathrm{values} = \sqrt{\sigma^2_\mathrm{bs} + \sigma^2_\mathrm{imp}}$.

\subsection{Practical advice for software users}\label{sec:advice}

We finish by giving a summary of the authors' approach to checking nested sampling calculations for challenging likelihoods where implementation errors may be present, based on our experience using nested sampling software.

We advise performing multiple nested sampling runs, and plotting the results to first assess their variation by eye as described in \Cref{sec:plots}.
One can then perform a rough check for implementation-specific effects using the techniques described in \Cref{sec:implementation_errors} and/or \Cref{sec:diagnostics}, depending on how many runs are available.
If implementation-specific errors are negligible:
\begin{itemize}
    \item Accuracy can be increased by simply calculating more runs and/or increasing the number of live points.
    \item The computational cost of future runs can be reduced by reducing the computational effort spent decorrelating samples (for example halving \PolyChord's \numrepeats{}, doubling \MultiNest's \efr{} or changing the equivalent setting in the software package used). After large changes to the settings, the new results should be checked for implementation-specific effects.
    \item Uncertainties on the results can be calculated using standard nested sampling methods such as the bootstrap resampling of threads, which will be accurate in this case.
\end{itemize}
In contrast, if implementation-specific effects are significant or are the dominant source of error:
\begin{itemize}
    \item Results should be recalculated with more live points and/or using more computational effort decorrelating samples (for example doubling \PolyChord's \numrepeats{}, halving \MultiNest's \efr{} or changing the equivalent setting in the software used). If the calculation is already very computationally costly, increasing the number of live points is typically the best option as this will also reduce errors from the stochasticity of the nested sampling algorithm.
    \item There may be an additional systematic bias present in the results of evidence calculations. The mean calculated value for results using the new settings should be checked to see if it is significantly different to the mean result produced with the previous settings.
    \item The uncertainty on the combined results from the nested sampling runs can be roughly estimated from~\eqref{equ:unc_comb}.
\end{itemize}

\section{Application to Planck survey data}\label{sec:planck}

We now apply the tests introduced in this paper to astronomical data from the \textit{Planck} survey, which measures anisotropies in the cosmic microwave background (CMB).
A detailed description of the associated cosmology and the $\Lambda$CDM concordance model is beyond the current scope; for this we refer the reader to \citet{PlanckCollaboration2013}.

Given the $\Lambda$CDM concordance model, we can describe the Universe's cosmology using only six parameters.
Four of these are ``late-time'' parameters, governing the physics of the Universe during and after reionisation: the present-day values of the Hubble constant $H_0$, the baryonic and cold dark matter fractions $\Omega_b$ and $\Omega_c$, and the optical depth of the CMB $\tau$.
The remaining two parameters delineate the primordial Universe through the amplitude $A_\mathrm{s}$ and tilt $n_\mathrm{s}-1$ of the power spectrum of comoving curvature perturbations.
To aid with MCMC sampling techniques, \texttt{cosmomc} \citep{cosmomc} reparameterises the matter fractions as $\Omega_b h^2$ and $\Omega_c h^2$ in terms of the reduced Hubble constant $h$, defined by $H_0 = 100h\:\text{km/s/Mpc}$, and in place of the Hubble constant uses $100\theta_{MC}$ ($100\times$ the ratio of the approximate sound horizon to the angular diameter distance).
For more details about the parameters, see the first \textit{Planck} parameters paper \citep{PlanckCollaboration2013}.

Given a set of cosmological parameters, using a Boltzmann code such as \texttt{camb} \citep{camb}, one may compute theoretical CMB power spectra, which are then provided as inputs to cosmological likelihoods derived from CMB observations.
We use the \texttt{Plik\_lite}  TT likelihood detailed by \citet{PlanckCollaboration2015a} and the default \texttt{CosmoChord} priors \citep[see][for more information]{Handley2015a}; these were used in \citet{PlanckCollaboration2015}.
The likelihood introduces a single additional nuisance parameter for measurement calibration, increasing the dimensionality of the parameter space to seven.

\begin{figure}
\centering
\includegraphics{img/line_plot_legend.pdf}
\begin{subfigure}{0.49\linewidth}
    \centering
    \includegraphics{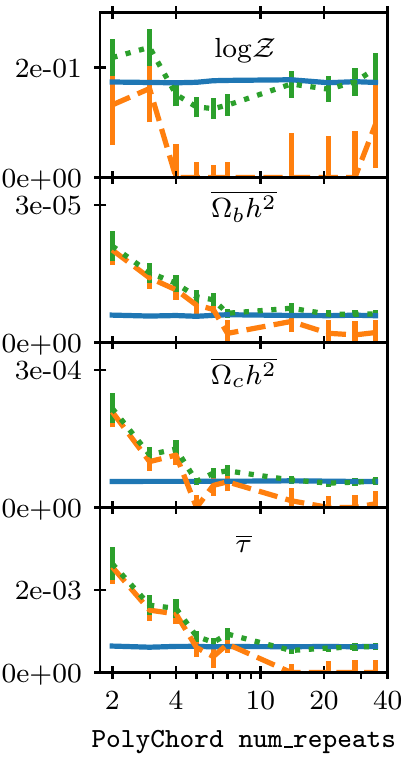}
\end{subfigure}\hfill
\begin{subfigure}{0.49\linewidth}
    \centering
    \includegraphics{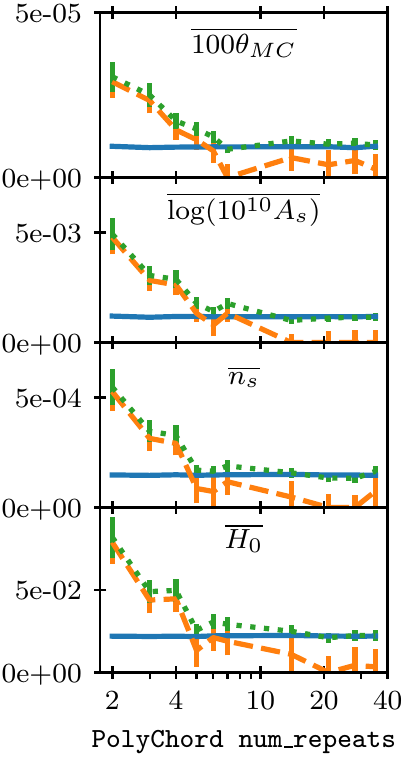}
\end{subfigure}
	\caption{Implementation-specific effects in calculations using \textit{Planck} data for different \PolyChord{} \numrepeats{} settings.
    The left column shows results for the evidence $\log \mathcal{Z}$, and the mean of the present day Baryon density $\Omega_b h^2$, present day cold matter density $\Omega_c h^2$ and Thompson scattering optical depth of the CMB $\tau$.
    The right column shows results for calculations of the mean of the ratio of the sound horizon to angular distance (scaled by 100) $100 \theta_{MC}$, the log power of the primordial curvature perturbations $\log(10^{10} A_s)$, the spectral index of the scalar primordial power spectrum $n_s$ and the present day Hubble constant (derived from the other parameters) $H_0$.
	Results for every $\numrepeats$ value were calculated using 25 runs, each with 500 live points.
	Blue solid lines show the mean bootstrap error estimate and orange dashed lines show implementation-specific effect estimates from~\eqref{equ:implementation_error}.
	Green dotted lines show the implementation-specific effects calculated using the root-mean-squared-error~\eqref{equ:implementation_error_rmse}; where the green dotted and orange dashed lines are equal, there is no systematic bias in the results.
	Error bars show the $1\sigma$ uncertainty on results for each \numrepeats{} value considered.
	}\label{fig:line_planck}
\end{figure}

\Cref{fig:line_planck} shows estimates of implementation-specific effects for calculations using the \textit{Planck} likelihoods and priors. Each calculation uses 500 live points.
As expected, there is a clear trend showing increasing \numrepeats{} reduces implementation-specific effects.
Furthermore in this case the \PolyChord{} setting $\numrepeats = 35$ (5 times the number of dimensions) is sufficient to make such effects small for all the calculations shown.

\begin{figure}
\centering
    \includegraphics{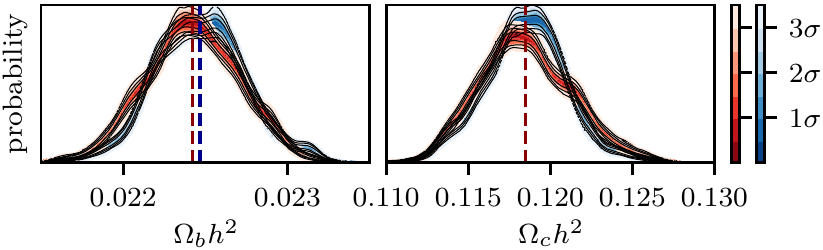}
    \includegraphics{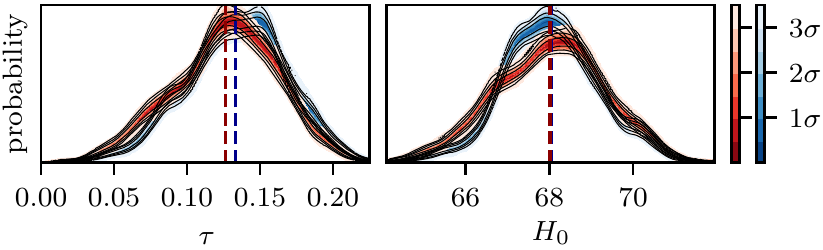}
    \caption{As for \Cref{fig:bs_param_dist} but using the \textit{Planck} survey likelihood.
    The first row shows the present day Baryon density $\Omega_b h^2$ and the present day cold matter density $\Omega_c h^2$; the second row shows the optical depth of the CMB $\tau$ and the present day Hubble constant $H_0$.
Each run uses 500 live points, and has $\numrepeats{}=1$ --- the low value is chosen to illustrate implementation-specific effects.
The coloured contours show iso-probability credible intervals on the marginalised posterior probability density function at each parameter value due to the stochasticity of the nested sampling algorithm.
The dashed dark blue and dark red lines show the estimated posterior means of each parameter for the blue and red runs respectively.}%
\label{fig:bs_param_planck}
\end{figure}

However, as in the test cases in previous sections, significant implementation-specifics are present in the calculations if \numrepeats{} is set too low.
This is illustrated in \Cref{fig:bs_param_planck} for $\numrepeats{}=1$; with this setting the two runs (in red and blue) differ by more than the uncertainty expected from the stochasticity of the nested sampling algorithm shown by the coloured distributions.
Such implementation-specific effects can also be detected with the diagnostic tests described in \Cref{sec:diagnostics} (we do not show these for brevity).
In addition, \Cref{fig:param_logx_planck} in Appendix~\ref{fig:param_logx_planck} shows a plot of the type described in \Cref{sec:param_logx_diagram} for the two runs in \Cref{fig:bs_param_planck}.

It should be noted that in cosmology one traditionally uses likelihoods with many more nuisance parameters than in this analysis. One of the innovations that \PolyChord{} provided to the \textit{Planck} collaboration was its ability to exploit a fast-slow hierarchy of parameter speeds \citep{cosmomc_fs}.
In this context, nuisance parameters that do not require recomputation of expensive parts of the likelihood may be varied at negligible cost in comparison with the slower cosmological parameters.
Increasing the number of steps in nuisance parameters directions greatly aids mixing and the reduction of implementation-specific errors.
However, a full analysis of this specific case is beyond the scope of this paper.

\section{Summary}

In this paper we introduced diagnostic tests for nested sampling software, which uses numerical techniques to generate approximately uncorrelated samples within hard likelihood constraints.
As a result additional errors may be produced which would not be present if the nested sampling algorithm was performed perfectly; we term these implementation-specific effects.
Detecting the presence of significant implementation-specific effects is of great importance for software users as it determines whether results and estimates of uncertainties can be relied upon, and if the settings should be changed.

We suggested two new diagnostic diagrams for visualising nested sampling results and uncertainties, and comparing runs; these are shown in \Cref{fig:bs_param_dist,sub:param_logx_gaussian,sub:param_logx_loggamma_mix,fig:bs_param_planck,fig:param_logx_planck}.
\Cref{sec:implementation_errors} introduced a quantitative measure of implementation-specific effects, which can be used to estimate them directly if enough runs are available to estimate the standard deviation of results.
In addition, \Cref{sec:diagnostics} provided two diagnostic tests which can be applied with only two runs.
The diagnostic tests and plots introduced in this paper are summarised in \Cref{tab:summary}.
We find that due to the larger errors from the stochasticity of the nested sampling algorithm in evidence calculations, implementation-specific errors form a smaller fraction of the total error in this case --- and are consequently less important and harder to detect than in parameter estimation.

\begin{table*}
\centering
\caption{Summary of the diagnostic tests and plots introduced in this paper.}\label{tab:summary}
\centering
\begin{tabular}{p{1.2in}p{0.6in}p{4.6in}}
\toprule
Diagnostic & Introduced & Summary \\
\midrule
Posterior distribution uncertainty plots
& \Cref{sec:bs_param_dist} &
Illustrates uncertainty on posterior distributions due to the stochasticity of the nested sampling algorithm.
Useful for comparing two or more runs to visually assess if their variation imples implementation-specific effects are present.
Examples are shown in \Cref{sub:bs_param_dist_gaussian,sub:bs_param_dist_loggamma_mix,fig:bs_param_planck}.
\\
$\log X$ plots
& \Cref{sec:param_logx_diagram} &
Shows the distribution of samples through the nested sampling process.
Can be used to understand and visualise posteriors and the manner in which the software explores them, as well as to assess if two runs are consistent.
Examples are shown in \Cref{sub:param_logx_gaussian,sub:param_logx_loggamma_mix,fig:param_logx_planck}.
\\
Calculating errors due to implementation-specific effects
& \Cref{sec:implementation_errors} &
Quantitatively estimates errors due to implementation-specific effects. This diagnostic provides the most information about the size implementation-specific effects, but it requires enough nested sampling runs to be able to estimate the standard deviation of their results.
\\
Testing correlations between threads
& \Cref{sec:thread_diagnostics} &
Checks if point estimates using threads from two runs are drawn from the same distribution. Can detect implementation-specific effects when only two runs are available, but does not give insight about their size.
\\
Testing sampling error distributions
& \Cref{sec:bootstrap_diagnostics} &
Checks if point estimates from different runs are consistent with each other given the stochasticity of the nested sampling algorithm. This can be done qualitatively with plots or quantitatively using statistical distances, and can be used when only two runs are available.
\\
\bottomrule
\end{tabular}
\end{table*}

In \Cref{sec:practical} we empirically tested the effects of software settings and the number of dimensions on implementation-specific effects, and discussed dealing with cases where nested sampling results are systematically biased.
The authors' practical advice for nested sampling software users based on our experience is summarised in \Cref{sec:advice}.
Finally, \Cref{sec:planck} demonstrated the application of our diagnostics to an astronomical problem using data from the \textit{Planck} survey.

We have written a publicly available software package \nestcheck{} \citep{Higson2018nestcheck}, which performs diagnostics on input nested sampling runs and produces plots like \Cref{fig:bs_param_dist,sub:param_logx_gaussian,sub:param_logx_loggamma_mix,fig:bs_param_planck,fig:param_logx_planck}; it can be downloaded at \nestcheckurl.

\section*{Acknowledgements}

We thank the anonymous reviewer for their detailed comments and suggestions.



\bibliographystyle{mnras}
\bibliography{library}

\begin{thebibliography}{}
\makeatletter
\relax
\def\mn@urlcharsother{\let\do\@makeother \do\$\do\&\do\#\do\^\do\_\do\%\do\~}
\def\mn@doi{\begingroup\mn@urlcharsother \@ifnextchar [ {\mn@doi@}
  {\mn@doi@[]}}
\def\mn@doi@[#1]#2{\def\@tempa{#1}\ifx\@tempa\@empty \href
  {http://dx.doi.org/#2} {doi:#2}\else \href {http://dx.doi.org/#2} {#1}\fi
  \endgroup}
\def\mn@eprint#1#2{\mn@eprint@#1:#2::\@nil}
\def\mn@eprint@arXiv#1{\href {http://arxiv.org/abs/#1} {{\tt arXiv:#1}}}
\def\mn@eprint@dblp#1{\href {http://dblp.uni-trier.de/rec/bibtex/#1.xml}
  {dblp:#1}}
\def\mn@eprint@#1:#2:#3:#4\@nil{\def\@tempa {#1}\def\@tempb {#2}\def\@tempc
  {#3}\ifx \@tempc \@empty \let \@tempc \@tempb \let \@tempb \@tempa \fi \ifx
  \@tempb \@empty \def\@tempb {arXiv}\fi \@ifundefined
  {mn@eprint@\@tempb}{\@tempb:\@tempc}{\expandafter \expandafter \csname
  mn@eprint@\@tempb\endcsname \expandafter{\@tempc}}}

\bibitem[\protect\citeauthoryear{Ahn \& Fessler}{Ahn \&
  Fessler}{2003}]{Ahn2003}
Ahn S.,  Fessler J.,  2003, EECS Department, University of Michigan, pp~1--2

\bibitem[\protect\citeauthoryear{Allison \& Dunkley}{Allison \&
  Dunkley}{2014}]{Allison2014}
Allison R.,  Dunkley J.,  2014, \mn@doi [Monthly Notices of the Royal
  Astronomical Society] {10.1093/mnras/stt2190}, 437, 3918

\bibitem[\protect\citeauthoryear{Beaujean \& Caldwell}{Beaujean \&
  Caldwell}{2013}]{Beaujean2013}
Beaujean F.,  Caldwell A.,  2013, arXiv preprint arXiv:1304.7808

\bibitem[\protect\citeauthoryear{Buchner}{Buchner}{2016}]{Buchner2016a}
Buchner J.,  2016, \mn@doi [Statistics and Computing]
  {10.1007/s11222-014-9512-y}, 26, 383

\bibitem[\protect\citeauthoryear{Chua, Hee, Handley, Higson, Moore, Gair,
  Hobson  \& Lasenby}{Chua et~al.}{2018}]{Chua2018}
Chua A. J.~K.,  Hee S.,  Handley W.~J.,  Higson E.,  Moore C.~J.,  Gair J.~R.,
  Hobson M.~P.,   Lasenby A.~N.,  2018, \mn@doi [Monthly Notices of the Royal
  Astronomical Society] {10.1093/mnras/sty1079}, 478, 28

\bibitem[\protect\citeauthoryear{Cowles \& Carlin}{Cowles \&
  Carlin}{1996}]{Cowles1996}
Cowles M.~K.,  Carlin B.~P.,  1996, Journal of the American Statistical
  Association, 91, 883

\bibitem[\protect\citeauthoryear{{DES Collaboration}}{{DES
  Collaboration}}{2018}]{DESCollaboration2017}
{DES Collaboration} 2018, \mn@doi [Physical Review D]
  {10.1103/PhysRevD.98.043526}, 98

\bibitem[\protect\citeauthoryear{Desvignes et~al.,}{Desvignes
  et~al.}{2016}]{Desvignes2016}
Desvignes G.,  et~al., 2016, \mn@doi [Monthly Notices of the Royal Astronomical
  Society] {10.1093/mnras/stw483}, 458, 3341

\bibitem[\protect\citeauthoryear{Fasano \& Franceschini}{Fasano \&
  Franceschini}{1987}]{Fasano1987}
Fasano G.,  Franceschini A.,  1987, \mn@doi [Monthly Notices of the Royal
  Astronomical Society] {10.1007/s10342-011-0499-z}, 225, 155

\bibitem[\protect\citeauthoryear{Feroz \& Hobson}{Feroz \&
  Hobson}{2008}]{Feroz2008}
Feroz F.,  Hobson M.~P.,  2008, \mn@doi [Monthly Notices of the Royal
  Astronomical Society] {10.1111/j.1365-2966.2007.12353.x}, 384, 449

\bibitem[\protect\citeauthoryear{Feroz, Hobson  \& Bridges}{Feroz
  et~al.}{2008}]{Feroz2009}
Feroz F.,  Hobson M.~P.,   Bridges M.,  2008, \mn@doi [Monthly Notices of the
  Royal Astronomical Society] {10.1111/j.1365-2966.2009.14548.x}, 398, 1601

\bibitem[\protect\citeauthoryear{Feroz, Hobson, Cameron  \& Pettitt}{Feroz
  et~al.}{2013}]{Feroz2013}
Feroz F.,  Hobson M.~P.,  Cameron E.,   Pettitt A.~N.,  2013, arXiv preprint
  arXiv:1306.2144

\bibitem[\protect\citeauthoryear{Handley}{Handley}{2018}]{Handley2018fgivenx}
Handley W.,  2018, \mn@doi [Journal of Open Source Software]
  {10.21105/joss.00849}, 3, 849

\bibitem[\protect\citeauthoryear{Handley, Hobson  \& Lasenby}{Handley
  et~al.}{2015a}]{Handley2015b}
Handley W.,  Hobson M.,   Lasenby A.,  2015a, \mn@doi [Monthly Notices of the
  Royal Astronomical Society] {10.1093/mnras/stv1911}, 15, 1

\bibitem[\protect\citeauthoryear{Handley, Hobson  \& Lasenby}{Handley
  et~al.}{2015b}]{Handley2015a}
Handley W.,  Hobson M.,   Lasenby A.,  2015b, \mn@doi [Monthly Notices of the
  Royal Astronomical Society: Letters] {10.1093/mnrasl/slv047}, 450, L61

\bibitem[\protect\citeauthoryear{Higson}{Higson}{2018a}]{Higson2018nestcheck}
Higson E.,  2018a, \mn@doi [Journal of Open Source Software]
  {10.21105/joss.00916}, 3, 916

\bibitem[\protect\citeauthoryear{Higson}{Higson}{2018b}]{Higson2018dypolychord}
Higson E.,  2018b, \mn@doi [Journal of Open Source Software]
  {10.21105/joss.00965}, 3, 965

\bibitem[\protect\citeauthoryear{Higson}{Higson}{2018c}]{Higson2018perfectns}
Higson E.,  2018c, \mn@doi [Journal of Open Source Software]
  {10.21105/joss.00985}, 3, 985

\bibitem[\protect\citeauthoryear{Higson, Handley, Hobson  \& Lasenby}{Higson
  et~al.}{2017}]{Higson2017b}
Higson E.,  Handley W.,  Hobson M.,   Lasenby A.,  2017, arXiv preprint
  arXiv:1704.03459

\bibitem[\protect\citeauthoryear{Higson, Handley, Hobson  \& Lasenby}{Higson
  et~al.}{2018}]{Higson2017a}
Higson E.,  Handley W.,  Hobson M.,   Lasenby A.,  2018, \mn@doi [Bayesian
  Analysis] {doi:10.1214/17-BA1075}, 13, 873

\bibitem[\protect\citeauthoryear{Hogg \& Foreman-Mackey}{Hogg \&
  Foreman-Mackey}{2018}]{Hogg2017}
Hogg D.~W.,  Foreman-Mackey D.,  2018, The Astrophysical Journal Supplement
  Series, 236, 11

\bibitem[\protect\citeauthoryear{Holm}{Holm}{1979}]{Holm1979}
Holm S.,  1979, \mn@doi [Scandinavian Journal of Statistics] {10.2307/4615733},
  6, 65

\bibitem[\protect\citeauthoryear{Joudaki et~al.,}{Joudaki
  et~al.}{2016}]{Joudaki2016}
Joudaki S.,  et~al., 2016, \mn@doi [Monthly Notices of the Royal Astronomical
  Society] {10.1093/mnras/stw2665}, 2052, 2033

\bibitem[\protect\citeauthoryear{Keeton}{Keeton}{2011}]{Keeton2011}
Keeton C.~R.,  2011, \mn@doi [Monthly Notices of the Royal Astronomical
  Society] {10.1111/j.1365-2966.2011.18474.x}, 414, 1418

\bibitem[\protect\citeauthoryear{Lewis}{Lewis}{2013}]{cosmomc_fs}
Lewis A.,  2013, \mn@doi [Physical Review D] {10.1103/PhysRevD.87.103529}, 87,
  103529

\bibitem[\protect\citeauthoryear{Lewis}{Lewis}{2015}]{Lewis2015}
Lewis A.,  2015, {GetDist: Kernel Density Estimation}

\bibitem[\protect\citeauthoryear{Lewis \& Bridle}{Lewis \&
  Bridle}{2002}]{cosmomc}
Lewis A.,  Bridle S.,  2002, \mn@doi [Physical Review D]
  {10.1103/PhysRevD.66.103511}, 66, 103511

\bibitem[\protect\citeauthoryear{Lewis, Challinor  \& Lasenby}{Lewis
  et~al.}{2000}]{camb}
Lewis A.,  Challinor A.,   Lasenby A.,  2000, \mn@doi [The Astrophysical
  Journal] {10.1086/309179}, 538, 473

\bibitem[\protect\citeauthoryear{Massey}{Massey}{1951}]{Massey1951}
Massey F.~J.,  1951, \mn@doi [Journal of the American Statistical Association]
  {10.1080/01621459.1951.10500769}, 46, 68

\bibitem[\protect\citeauthoryear{Murray}{Murray}{2007}]{Murray2007}
Murray I.,  2007, PhD thesis, University College London

\bibitem[\protect\citeauthoryear{{Planck Collaboration}}{{Planck
  Collaboration}}{2013}]{PlanckCollaboration2013}
{Planck Collaboration} 2013, \mn@doi [Astronomy {\&} Astrophysics]
  {10.1051/0004-6361/201321591}, 571, 1

\bibitem[\protect\citeauthoryear{{Planck Collaboration}}{{Planck
  Collaboration}}{2016a}]{PlanckCollaboration2015a}
{Planck Collaboration} 2016a, \mn@doi [Astronomy and Astrophysics]
  {10.1051/0004-6361/201526926}, 594, A11

\bibitem[\protect\citeauthoryear{{Planck Collaboration}}{{Planck
  Collaboration}}{2016b}]{PlanckCollaboration2015}
{Planck Collaboration} 2016b, \mn@doi [Astronomy {\&} Astrophysics]
  {10.1051/0004-6361/201525898}, 594, A20

\bibitem[\protect\citeauthoryear{Samushia et~al.,}{Samushia
  et~al.}{2014}]{Samushia2014}
Samushia L.,  et~al., 2014, \mn@doi [Monthly Notices of the Royal Astronomical
  Society] {10.1093/mnras/stu197}, 439, 3504

\bibitem[\protect\citeauthoryear{Scholz \& Stephens}{Scholz \&
  Stephens}{1987}]{Scholz1987}
Scholz F.~W.,  Stephens M.~A.,  1987, \mn@doi [Journal of the American
  Statistical Association] {10.1080/01621459.1987.10478517}, 82, 918

\bibitem[\protect\citeauthoryear{Scott}{Scott}{2015}]{Scott2015}
Scott D.~W.,  2015, {Multivariate Density Estimation: Theory, Practice, and
  Visualization}.
John Wiley {\&} Sons

\bibitem[\protect\citeauthoryear{Skilling}{Skilling}{2006}]{Skilling2006}
Skilling J.,  2006, \mn@doi [Bayesian Analysis] {10.1214/06-BA127}, 1, 833

\bibitem[\protect\citeauthoryear{Wolpert \& Macready}{Wolpert \&
  Macready}{1997}]{Wolpert1997}
Wolpert D.~H.,  Macready W.~G.,  1997, \mn@doi [IEEE Transactions on
  Evolutionary Computation] {10.1109/4235.585893}, 1, 67

\makeatother
\end{thebibliography}



\appendix

\section{Code}

The code used to perform the numerical tests and generate the results in this paper can be downloaded at \diagnosticcodeurl; this provides examples of \nestcheck{}'s use.

\section{Numerical results tables}\label{app:tables}

\Cref{tab:gaussian,tab:loggamma_mix} given numerical results for the nested sampling runs plotted in \Cref{fig:imp_error_bars}.

\begin{table}
\centering
\caption{Calculation error results for the 100 nested sampling runs with a Gaussian likelihood shown in \Cref{fig:imp_error_bars}.
The first two rows shows the true value for each estimator and the mean calculation result.
The next three rows show the bootstrap error estimate, implementation error estimate~\eqref{equ:implementation_error} and the ratio of the implementation estimate to the standard deviation of results.
The final three rows show the root-mean-squared-error, the implementation-specific effects estimate from~\eqref{equ:implementation_error_rmse}, and the ratio of the two.
Columns show results for the log-evidence and the mean of the first three parameters.
Numbers in parentheses show the $1\sigma$ numerical uncertainty on the final digit.}\label{tab:gaussian}
\begin{tabular}{lllll}
\toprule
{} & $\log \mathcal{Z}$ & $\thmean{1}$ & $\thmean{2}$ & $\thmean{3}$ \\
\midrule
True Value              &                   -40.9434 &                        0.0000 &                        0.0000 &                        0.0000 \\
Mean Result              &                  -40.93(3) &                      0.002(2) &   0.000(2)&                     0.000(2) \\
$\sigma_\mathrm{values}$               &                    0.33(2) &                      0.022(2) &                      0.019(1) &                      0.019(1) \\
$\sigma_\mathrm{bs}$       &                   0.326(3) &                     0.0223(2) &                     0.0223(2) &                     0.0221(2) \\
$\sigma_\mathrm{imp}$       &                   0.07(11) &                      0.000(7) &                      0.000(3) &                      0.000(3) \\
$\sigma_\mathrm{imp} / \sigma_\mathrm{values}$  &                   0.20(33) &                      0.00(34) &                      0.00(17) &                      0.00(17) \\
Values RMSE              &                    0.33(2) &                      0.022(2) &                      0.019(1) &                      0.019(1) \\
$\sigma_\mathrm{imp,RMSE}$      &                   0.06(11) &                      0.000(7) &                      0.000(2) &                      0.000(3) \\
$\sigma_\mathrm{imp,RMSE} / \mathrm{RMSE}$ &                   0.17(33) &                      0.00(34) &                      0.00(17) &                      0.00(19) \\
\bottomrule
\end{tabular}
\end{table}

\begin{table}
\centering
\caption{As in \Cref{tab:gaussian} but for calculations using the LogGamma mix likelihood~\eqref{equ:loggamma_mix}.}\label{tab:loggamma_mix}
\begin{tabular}{lllll}
\toprule
{} & $\log \mathcal{Z}$ & $\thmean{1}$ & $\thmean{2}$ & $\thmean{3}$ \\
\midrule
True Value              &                   -40.9434 &                       -0.5772 &                        0.0000 &                       -0.5772 \\
Mean Result              &                  -40.84(3) &                     -0.49(18) &                     -0.22(18) &                     -0.572(3) \\
$\sigma_\mathrm{values}$               &                    0.34(2) &                      1.78(13) &                      1.81(13) &                      0.032(2) \\
Values RMSE              &                    0.36(2) &                      1.77(12) &                      1.81(10) &                      0.032(2) \\
$\sigma_\mathrm{bs}$       &                   0.309(3) &                      0.217(2) &                      0.215(2) &                     0.0300(3) \\
$\sigma_\mathrm{imp}$       &                    0.15(8) &                      1.76(13) &                      1.80(13) &                       0.01(1) \\
$\sigma_\mathrm{imp} / \sigma_\mathrm{values}$  &                   0.43(23) &                      0.993(1) &                      0.993(1) &                      0.31(30) \\
$\sigma_\mathrm{imp,RMSE}$      &                    0.18(6) &                      1.76(13) &                      1.80(10) &                      0.011(9) \\
$\sigma_\mathrm{imp,RMSE} / \mathrm{RMSE}$ &                   0.50(14) &                      0.992(1) &                     0.9930(8) &                      0.33(28) \\
\bottomrule
\end{tabular}
\end{table}

\section{Planck survey data $\log X$ plot}\label{app:planck_logx_plot}

\begin{figure}
\centering
    \includegraphics{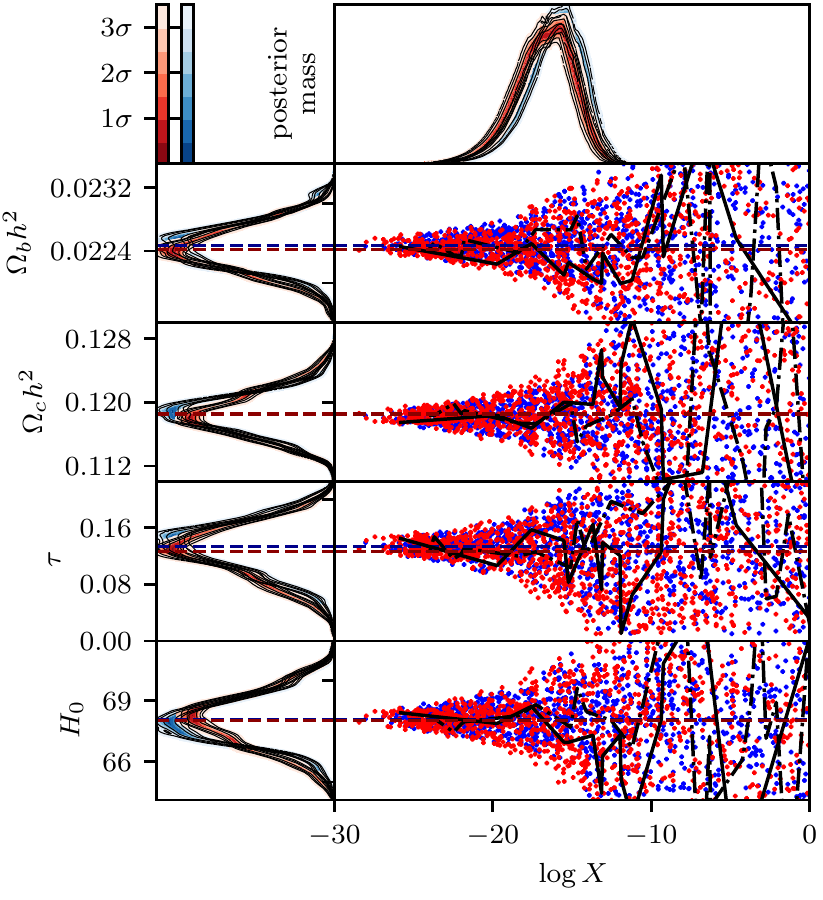}
    \caption{As for \Cref{sub:param_logx_loggamma_mix} but using the \textit{Planck} survey likelihood.
The two runs (shown in red and blue) are the same ones used for \Cref{fig:bs_param_planck}.
The top right panel shows the relative posterior mass (total weight assigned to all samples in that region) as a function of $\log X$.
The final 4 rows show the present day Baryon density $\Omega_b h^2$, the present day cold matter density $\Omega_c h^2$, the optical depth of the CMB $\tau$ and the present day Hubble constant $H_0$.
The coloured contours show iso-probability credible intervals on the marginalised posterior probability density function at each parameter or $\log X$ value.
In each row, the estimated posterior means for the blue and red runs are shown with dashed dark blue and dark red lines.
The solid and dot dash black lines show the evolution of an individual thread chosen at random from the red and blue runs respectively.}%
\label{fig:param_logx_planck}
\end{figure}

\Cref{fig:param_logx_planck} shows a plot of samples' distributions in $\log X$ (of the type described in \Cref{sec:param_logx_diagram}) using the same runs as \Cref{fig:bs_param_planck}.
In this case as the posterior is relatively simple and unimodal, and the samples overlap closely.

\bsp{}
\label{lastpage}
\end{document}